\documentclass[letterpaper,twocolumn,dvipsnames,10pt]{article}
\usepackage{usenix}
\usepackage[available,functional,reproduced]{usenixbadges}

\usepackage{float}
\usepackage{graphicx}
\usepackage{amsmath}
\usepackage{amssymb}
\usepackage{amsfonts}
\usepackage{balance}
\usepackage{booktabs}
\PassOptionsToPackage{hyphens}{url}\usepackage{hyperref}
\usepackage[dvipsnames]{xcolor}
\usepackage{xspace}
\usepackage{tabularx}
\usepackage{algorithm}
\usepackage[noend]{algpseudocode}
\usepackage{caption}
\usepackage{subcaption}
\usepackage{overpic}
\usepackage{xspace}
\usepackage{pifont}
\usepackage{listings}
\usepackage{multirow}
\usepackage{makecell}
\usepackage{mdframed}
\usepackage[section]{placeins}
\usepackage[compact]{titlesec}
\usepackage[scaled=.85]{FiraMono}
\usepackage[T1]{fontenc}

\usepackage{enumitem}
\setitemize{noitemsep,topsep=0pt,parsep=0pt,partopsep=0pt}

\newcommand{\captionfonts}{\small}
\makeatletter  
\long\def\@makecaption#1#2{%
  \vskip 0.1in
  \sbox\@tempboxa{{\captionfonts \textbf{#1}. #2}}%
  \ifdim \wd\@tempboxa >\hsize
    {\captionfonts \textbf{#1}. #2\par}
  \else
    \hbox to\hsize{\hfil\box\@tempboxa\hfil}%
  \fi
  \vskip 0in}
\makeatother   

\def\systemnameRaw{ServeGen}

\def\sysname{\systemnameRaw\xspace}

\newcounter{findingno}
\DeclareRobustCommand{\finding}[1]{%
    \refstepcounter{findingno}%
    \thefindingno\label{#1}}

\def\qmax{{\texttt{M-large}}\xspace}
\def\qturbo{{\texttt{M-small}}\xspace}
\def\qplus{{\texttt{M-mid}}\xspace}
\def\qlong{{\texttt{M-long}}\xspace}
\def\spark{{\texttt{M-rp}}\xspace}
\def\lingma{{\texttt{M-code}}\xspace}

\def\image{{\texttt{mm-image}}\xspace}
\def\audio{{\texttt{mm-audio}}\xspace}
\def\video{{\texttt{mm-video}}\xspace}
\def\omni{{\texttt{mm-omni}}\xspace}

\def\ds{{\texttt{deepseek-r1}}\xspace}
\def\dsqwen{{\texttt{deepqwen-r1}}\xspace}

\newif\ifshowcomment
\showcommenttrue

\ifshowcomment
\newcommand{\todo}[1]{\textsf{\color{red}{todo: {#1}}}}

\newcommand{\ray}[1]{\textsf{\color{orange}{[Ray: {#1}]}}}

\else
\newcommand{\todo}[1]{}
\newcommand{\ray}[1]{}
\fi

\newcommand{\stitle}[1]{\vspace{1ex}\noindent{\bf #1}}

\newcommand{\ie}{\emph{i.e.,}\xspace}
\newcommand{\eg}{\emph{e.g.,}\xspace}

\newcommand{\etc}{\emph{etc.}\xspace}

\begin{document}
\pagestyle{empty}

\title{
    \Large \bf
    \sysname: Workload Characterization and Generation of
    Large Language Model Serving in Production
}

\author{
    \rm{
        Yuxing Xiang$^{\text{1,2}}$ \qquad
        Xue Li$^{\text{2}}$ \qquad
        Kun Qian$^{\text{2}}$ \qquad
        Yan Zhang$^{\text{1}}$ \qquad
    }\\
    \rm{
        Wenyuan Yu$^{\text{2}}$ \qquad
        Ennan Zhai$^{\text{2}}$ \qquad
        Xin Jin$^{\text{1}}$ \qquad
        Jingren Zhou$^{\text{2}}$ \qquad
    }
    \\
    {\small $^{\text{1}}$School of Computer Science, Peking University \qquad $^{\text{2}}$Alibaba Group}
}

\maketitle

\noindent\textbf{Abstract.} With the widespread adoption of Large Language Models (LLMs), serving LLM inference requests 
has become an increasingly important task, attracting active research advancements.
Practical workloads play an essential role in this process: they are critical for motivating 
and benchmarking serving techniques and systems.
However, the existing understanding of real-world LLM serving workloads is limited due to 
the lack of a comprehensive workload characterization. Prior analyses remain 
insufficient in scale and scope, thus failing to fully capture intricate workload characteristics.

In this paper, we fill the gap with an in-depth characterization of LLM serving workloads 
collected from our worldwide cloud LLM serving service, covering not only language models 
but also emerging \emph{multimodal} and \emph{reasoning} models, unveiling important
\emph{new} findings in each case.
Moreover, based on our findings, we propose \sysname, a principled framework for 
generating realistic LLM serving workloads by composing them on a per-client basis. 
Practical use cases validate that \sysname achieves more accurate performance 
benchmarking compared to naive workload generation, and reveals new design 
implications that could otherwise be overlooked.
\sysname is open-sourced at \url{https://github.com/alibaba/ServeGen}.
\vspace{1em}
\section{Introduction}
\label{sec:introduction}

In recent years, the rapid evolution of Large Language Models (LLMs)~\cite{deepseek-v3, qwen2.5, gpt-4o, grok-3} has enabled fundamentally new applications, 
with large-scale deployment in production clusters serving substantial user traffic every day~\cite{deepseek-inference}.
To accommodate this growing demand, a large body of research has focused on optimizing LLM serving 
in terms of model serving latency~\cite{fastserve,fastgen}, resource utilization~\cite{vllm,SarathiServe,splitwise}, 
service quality~\cite{distserve,liu2024andes}, and beyond~\cite{sglang,ServerlessLLM,wu2024loongserve}.

Inference serving workloads play an important role in this innovation process:
they motivate the design of new optimization techniques and systems, and the
effectiveness of the latter must be validated under respective workloads. 
Yet, there is an absence of comprehensive, production-scale characterization 
of real-world serving workloads. The status quo is a mixture of 
$(i)$ adapted workloads from general deep-learning or cloud computing tasks~\cite{Philly,Helios,Acme,PAI,MAF} 
(\eg using function invocations in serverless workloads as inference requests), and 
$(ii)$ optimization-~\cite{splitwise,cao2025locality,mooncake,DynamoLLM} or
pattern-specific~\cite{burstgpt,LMM} analyses (\eg detailing only certain patterns), 
which remain insufficient in scale and scope.

The lack of practical workload characterization poses two obstacles to the 
innovation process of LLM serving systems.
First, the many \emph{uncharacterized} aspects of real-world workloads hinder new insights 
and motivations, especially for emerging scenarios such as serving multimodal~\cite{Qwen2.5-Omni,chen2025janus} 
and reasoning~\cite{deepseek-r1,gpt-o1} models.
Second, even for serving normal (\ie non-reasoning) language models that have been 
extensively studied, the inadequate understanding of real-world workloads 
may still result in \emph{unrealistic} benchmarking when evaluating emerging optimizations.
The \emph{de facto} approach (referred to as \textsc{Naive}) adopted by many studies~\cite{fastserve,shepherd,AlpaServe,wu2024loongserve} 
generates workloads by simply combining certain arrival \emph{traces}
(\eg sampled from Poisson or Gamma processes, or scaled from published traces~\cite{MAF}) 
with \emph{datasets} (\eg ShareGPT~\cite{sharegpt}).\footnote{
    Prior work has used the terms ``trace'', ``dataset'', and 
    ``workload'' interchangeably. 
    In our discussion, ``trace'' denotes request arrival timestamps, while ``dataset''
    refers to request data distributions exclusively.
}
However, prior experience in cloud workload modeling~\cite{verma2014evaluating,tracegen} 
has highlighted more intricate workload patterns, such as ``heterogeneity''~\cite{dynamics} and 
``imbalance''~\cite{imbalance}, revealing that ``naively-generated workloads are misleadingly easier to 
serve than real historial ones''~\cite{tracegen}. In practice, scaling serving  
optimizations to deployment has been met with unforeseen difficulties,  
such as performance degradation~\cite{pdserve} and major revisions in system design~\cite{vllm-pd}.

\begin{table*}[t]
    \centering
    \caption{
        The list of workloads and models in our study. 
    }
    \label{tab:models}
    \small
    \begin{tabular}{llp{4.5cm}ll}
    \toprule
    \textbf{Category} & \textbf{Name} & \textbf{Model} & \textbf{Description} & \textbf{Workload Information} \\
    
    \midrule
    Language   & \qmax & General model (310B) & Largest, general-purpose & 240M requests (one month) \\
                & \qplus & General model (72B) & Balanced, general-purpose & 2.1B requests (one month) \\
                & \qturbo & General model (14B) & Cheapest, general-purpose & 767M requests (one month) \\
                & \qlong & General model (72B, 10M context) & Long-document comprehension & 48M requests (one week) \\
                & \spark & Domain-specific model & Role-playing & 49M requests (one week) \\
                & \lingma & Domain-specific model & Code completion & 276M requests (one week) \\
        
    \midrule
    Multimodal  & \image & Qwen2.5-VL-72B & Image \& text input & 28M requests (one month) \\
                & \audio & Qwen2-Audio-7B & Audio \& text input & 420K requests (one month) \\
                & \video & Qwen2.5-VL-72B & Video \& text input & 1.2M requests (one month) \\
                & \omni & Qwen2.5-Omni-7B & Omni-modal input & 8.7M requests (one week) \\
    
    \midrule
    Reasoning   & \ds & deepseek-r1-671B & Full reasoning model & 14.0M requests (one week) \\
                & \dsqwen & deepseek-r1-distill-qwen-32B  & Distilled reasoning model &  4.8M requests (one week)\\
    \bottomrule
    \end{tabular}
    \vspace{-2em}
\end{table*}

As a large cloud inference service provider, we aim to fill this gap with an
extensive and detailed characterization of real-world LLM serving workloads, 
analyzing a diverse range of models (see Table~\ref{tab:models}) and billions of requests 
collected from our production clusters over four months.
We provide a comprehensive analysis of LLM serving workloads
that covers language (\S\ref{sec:text}), multimodal (\S\ref{sec:mm}), and reasoning (\S\ref{sec:reasoning}) 
workloads, unveiling important \emph{new} findings.
We release a principled framework, \sysname, which allows practitioners to incorporate our 
findings and generate realistic workloads that better reflect system performance 
compared to \textsc{Naive} workload generation (\S\ref{sec:reconstruct}), 
thus facilitating the motivation and evaluation of ongoing research.

\stitle{Characterizing language model workloads.}
We begin with a characterization of various (non-reasoning) language model workloads based on their arrival 
patterns~(\S\ref{sec:text:arrival}) and input/output lengths~(\S\ref{sec:text:io}).
While there is prior work analyzing language model workloads, our analysis 
yields important new findings:
$(i)$ request arrivals exhibit a complex bursty pattern that goes beyond any single stochastic process (\eg 
a gamma process is not necessarily the best fit in all cases); and
$(ii)$ the input/output length distributions can be modeled by combinations of classic distributions, 
but the corresponding parameters vary significantly over time.
Considering these findings, we conduct a deep-dive analysis by decomposing 
the workloads by clients~(\S\ref{sec:text:client}), revealing a \emph{causal modeling} of real-world workloads: 
most nondeterministic patterns in request arrivals (\eg bursts) and length distributions 
(\eg high dynamics over time) are caused by several top clients, 
while the behaviors of most clients remain stable and predictable.
This finding is valuable for generating realistic workloads.

\stitle{Characterizing multimodal and reasoning workloads.}
We also analyze inference serving workloads of emerging multimodal and reasoning models, 
highlighting their unique characteristics.
For multimodal workloads, we report significant load variance across modalities (\S\ref{sec:mm:data}) 
and substantial request heterogeneity (\S\ref{sec:mm:hetero}), unveiling 
inefficiencies in the prefill phase of LLM inference.
For reasoning workloads, the long and bimodal distribution of reasoning lengths (\S\ref{sec:reasoning:data}) 
and the more stable arrival pattern from multi-turn conversations (\S\ref{sec:reasoning:multi-turn}) 
present both challenges and opportunities in optimizing the decoding phase.
In both scenarios, similarly, we analyze the workloads with client decomposition (\S\ref{sec:mm:client} and \S\ref{sec:reasoning:client}) 
to deepen our characterization, again capturing the diverse patterns through causal modeling.

\stitle{Workload generation.}
While the aforementioned findings help motivate the design of next-generation LLM serving systems, 
it remains crucial for practitioners to be able to evaluate said systems with realistic 
workloads. 
However, full-scale production workloads are not always available due to privacy concerns, 
particularly in emerging serving scenarios.
Moreover, the few published workloads~\cite{burstgpt,splitwise,mooncake,cao2025locality,DynamoLLM} 
are limited to a specific scale and are subject to the so-called ``workload churn''~\cite{churn}.
To better share our insights and further facilitate the community, 
we build and release \sysname, a workload generation framework to generate realistic
serving workloads.
\sysname performs principled modeling of workloads on a \emph{per-client} basis based on our 
findings to generate realistic workloads, and is easy to use~(\S\ref{sec:reconstruct:framework}). 
Our evaluation shows that \sysname outperforms the \textsc{Naive} workload generation approach by 
producing workloads that better align with real ones (\S\ref{sec:reconstruct:accuracy}).
Additionally, we demonstrate that \sysname is beneficial for benchmarking and motivating 
better serving system designs in production via case studies
on instance provisioning and PD-disaggregation~\cite{distserve}.

\stitle{Contributions.}
Our main contributions are as follows.

\begin{itemize}[leftmargin=*]
    \setlength\itemsep{-0em}
    \item We provide a comprehensive study of real-world LLM serving workloads
    in a large-scale production environment, which not only covers 
    language models, but also emerging multimodal and reasoning models.

    \item We characterize production-level LLM serving workloads
    and conduct in-depth analysis by client decomposition,
    revealing important new findings.

    \item We release \sysname, a principled framework for
    generating realistic serving workloads based on our findings
    to help motivate and benchmark future research.
\end{itemize}
\smallskip
\section{Background}
\label{sec:background}

\subsection{LLM Basics}
\label{sec:background:llm}

\stitle{Basic LLM inference.}
The typical inference workflow for an LLM request comprises two key phases: 
\emph{prefill} and \emph{decoding}. 
In the prefill phase, all \emph{input} tokens in the user prompt are processed to 
generate the first \emph{output} token. Subsequently, the decoding phase auto-regressively 
generates the rest of the output tokens sequentially, until either the generation of an end-of-sequence 
(EOS) token or a predefined maximum output length is reached. 
In both phases, requests are commonly batched~\cite{orca} and processed simultaneously to enhance  
the serving throughput.
Consequently, the arrival pattern and input/output lengths of requests are strongly 
relevant to LLM inference performance, as they impact the batching result and computational 
load during execution.

\stitle{Multimodal models.}
Multimodal LLMs~\cite{Qwen2.5-Omni,chen2025janus} are extended with the ability to process and 
integrate multiple types of data beyond text prompts, such as images, audio, and video, 
allowing for richer user interactions.
In a typical multimodal inference workflow, a model must first process its multimodal inputs 
through a series of \emph{downloading} (fetching data from URLs), \emph{normalizing} (\eg resizing 
images or resampling audio), and \emph{encoding} (through modality-specific adapters, such as ViT~\cite{vit})
stages to obtain their embeddings, which are fused with the text embeddings. 
The inference then proceeds in a process identical to basic LLM serving. 
As such, multimodal data distributions play a crucial role in the 
inference performance of multimodal LLMs.

\stitle{Reasoning models.}
A significant recent progress in LLMs is the rise of reasoning~\cite{gpt-o1,deepseek-r1}, 
which shows remarkable capabilities in complex coding, math, and problem-solving tasks. 
These models' output tokens are divided into two sections: first the \emph{reason} tokens, 
where the model performs test-time computation~\cite{TTC}, and second the \emph{answer} tokens 
that actually answer the input prompt. 
This behavior makes reasoning workloads stand out from normal language model workloads, 
altering the workload statistics (\eg longer outputs) while also potentially enabling  
new optimizations. 

\begin{table}[t]
    \centering
    \caption{
        Comparison between our work and prior characterizations of LLM serving workloads.
        Dashes indicate unavailable data. 
    }
    \label{tab:related}
    \footnotesize
    \begin{tabular}{lp{2cm}lll}
    \toprule
                & \textbf{Ours} & \textbf{BurstGPT}~\cite{burstgpt} & \textbf{LMM}~\cite{LMM}\\
    \midrule
    
    & \multicolumn{3}{c}{\textbf{Characterization $\triangleright$ Scale}}\\
    \midrule
    Duration    & 4 months  & 4 months  & 2 days \\
    \#Models     & 12 & 2  & - \\
    \#Requests  & 3.54B   & 5.29M  & - \\
    \midrule

    &  \multicolumn{3}{c}{\textbf{Characterization $\triangleright$ Scope}}\\
    \midrule
    Workloads   & Language, Multimodal, Reasoning & Language & Image-modal \\
    Patterns   & Variant burstiness        & Variant burstiness    & Image data    \\
                        & Distribution shifts    &                       & distribution  \\
                        & Conversations             &                       &   \\
    \midrule

    &  \multicolumn{3}{c}{\textbf{Workload Generation}}\\
    \midrule
    Approach    & Parameterized & Parameterized & \textsc{Naive} \\
                & clients       & burstiness &  \\

    \bottomrule
    \end{tabular}
    \vspace{-2.5em}
\end{table}

\subsection{LLM Serving Workloads}
\label{sec:background:workload}

\stitle{Workload characterization and generation.}
Optimization of LLM serving systems promises significant performance gains and substantial 
cost reductions. However, achieving this goal requires a deep understanding of real-world 
workloads, which is often unavailable due to the absence of a comprehensive 
production-scale workload characterization. 
Table~\ref{tab:related} summarizes related work on LLM inference workload 
characterization, 
omitting various brief analyses found in other work~\cite{splitwise,cao2025locality,mooncake,DynamoLLM} 
that are optimization-specific and more limited in scope. 
As shown by the comparison, state-of-the-art characterizations are lacking in
terms of scale, and leave many workload patterns \emph{uncharacterized}.
Furthermore, this inadequacy results in \emph{unrealistic} workload generation
approaches, restricting practitioners to workloads that
cannot fully capture real-world patterns.
Thus, we are motivated to perform a more extensive characterization 
of real LLM serving workloads in production.
We then share our insights by building and releasing \sysname, a principled framework 
for generating realistic serving workloads to foster future research. 

\stitle{Workload source.}
Alibaba Cloud Model Studio is a cutting-edge AI model service platform that enables users to build 
and use various custom model services.
It hosts over 200 foundation models and thousands of fine-tuned models, supports applications deployed 
by hundreds of enterprises, and serves millions of requests each day.
To support such a high and diverse workload, Model Studio maintains O(10K) GPUs distributed 
in dozens of regions and zones, making it a world-wide large model service platform.

Our characterization is supported by real inference workloads running in Model Studio. 
The analyzed workloads span four months, containing 12 models and billions of requests 
from datacenters in different geolocations, as shown in Table~\ref{tab:models}.
We source request metadata from our log store for the backend inference engines, 
collecting detailed information including request arrival and execution times, payload (\eg input and 
output lengths, chat histories, and multimodal inputs), and other relevant data, 
all sanitized to respect client privacy. 
Notably, these data are agnostic of serving system internals or implementation 
details (as supposed to metrics like request completion time), allowing for an unbiased 
and comprehensive understanding of LLM serving workloads.
\section{Characterizing Language Workloads}
\label{sec:text}

This section analyzes language model workloads listed in Table~\ref{tab:models}. 
We report findings for arrival times~(\S\ref{sec:text:arrival}) 
and input/output length distributions~(\S\ref{sec:text:io}), two essential traits that affect 
the performance of LLM serving systems. 
Importantly, we show that much of the complex underlying patterns behind our findings 
can be explained by client decomposition~(\S\ref{sec:text:client}).

\subsection{Request Arrival Pattern}
\label{sec:text:arrival}

\begin{figure}[t!]
    \centering
    \begin{subfigure}[t]{0.43\linewidth}
      \centering
      \includegraphics[width=\linewidth]{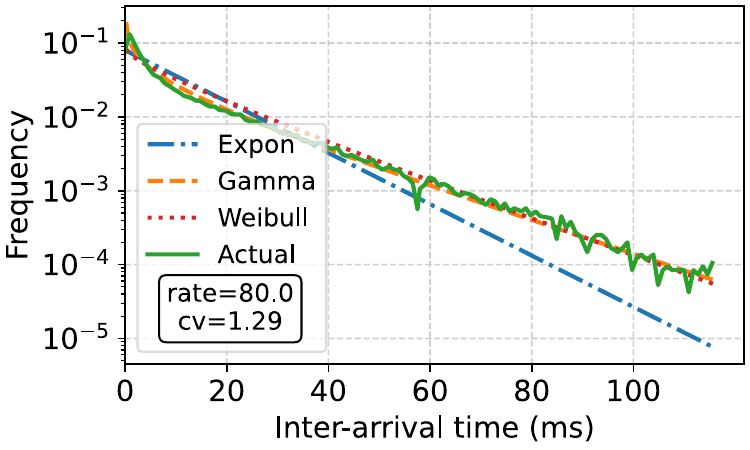}
      \vspace{-0.25in}
      \caption{\qmax}
      \label{fig:iat:1}
    \end{subfigure}
    \begin{subfigure}[t]{0.43\linewidth}
      \centering
      \includegraphics[width=\linewidth]{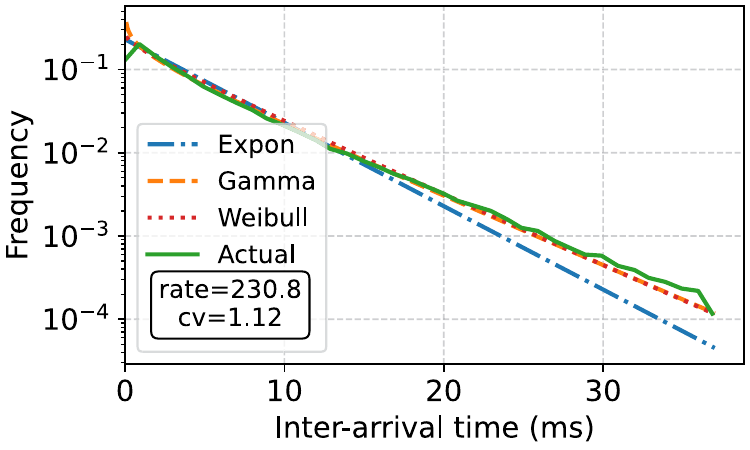}
      \vspace{-0.25in}
      \caption{\qturbo}
      \label{fig:iat:2}
    \end{subfigure}

    \centering
    \begin{subfigure}[t]{0.43\linewidth}
        \centering
        \includegraphics[width=\linewidth]{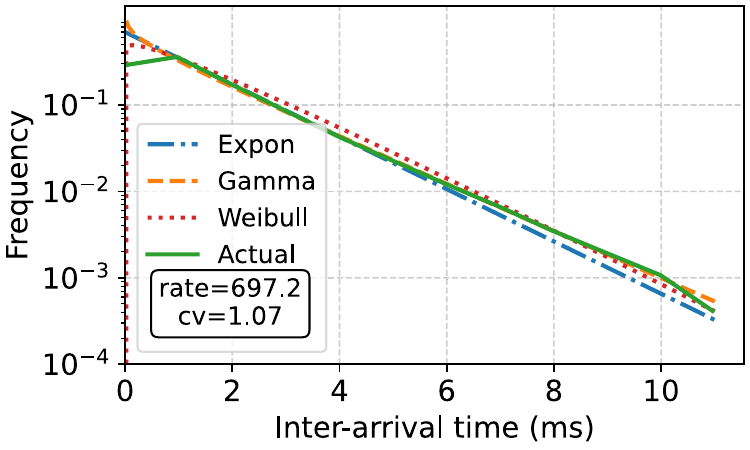}
        \vspace{-0.25in}
        \caption{\qplus}
        \label{fig:iat:3}
    \end{subfigure}
    \begin{subfigure}[t]{0.43\linewidth}
        \centering
        \includegraphics[width=\linewidth]{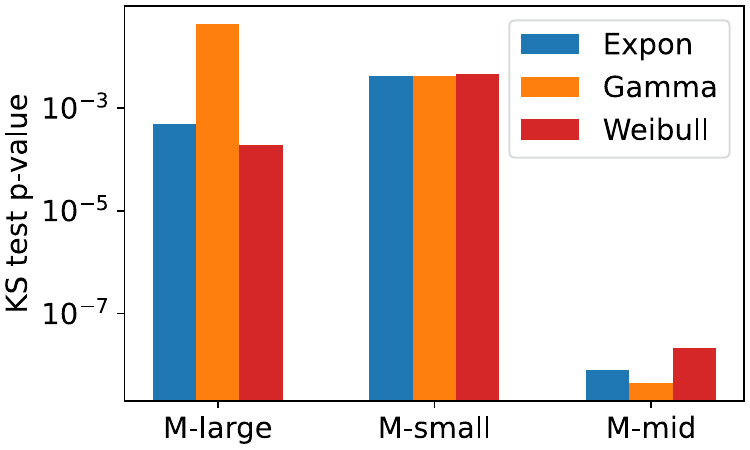}
        \vspace{-0.25in}
        \caption{Hypothesis test}
        \label{fig:iat:4}
    \end{subfigure}
    \vspace{-0.075in}
    \caption{
        Inter-arrival time characterization. 
    }
    \label{fig:iat}
    \vspace{-0.175in}
\end{figure}

\stitle{Bursty short-term arrival patterns.}
Figure~\ref{fig:iat} characterizes the inter-arrival time (IAT) distributions for \qmax, 
\qturbo, and \qplus within a 20-minute window.
Conforming to existing analyses~\cite{burstgpt,MAF}, we find that the arrival patterns 
exhibit notable \emph{burstiness}, indicated by coefficients of variation (CVs) greater than 1. 
Consequently, Poisson processes (CV=1) often poorly model bursty workloads (such as in Figure~\ref{fig:iat:1}), 
where Gamma and Weibull processes are better alternatives.
However, there is not a universally best modeling, which is validated in Figure~\ref{fig:iat:4}, 
where we apply the Kolmogorov-Smirnov (KS) test to check whether the IATs came from Exponential, 
Gamma, or Weibull distributions.\footnote{
    Indeed, these p-values are too small to deny the null hypothesis (that the arrival
    is modeled by some distribution), which is common for the KS test when the sample size is large. 
    However, comparing p-values remains helpful.
}
None of the distributions has the largest p-value consistently, 
indicating variable goodness of fit. In fact, the best-fit choices are different 
across workloads: Gamma for \qmax, Weibull for \qplus, and even Exponential 
can be a good fit for \qturbo.
Practically, this implies that arrival patterns in real-world workloads should 
be modeled flexibly using different distributions to better preserve their characteristics.

\begin{figure}[t!]
    \centering
    \includegraphics[width=0.92\linewidth]{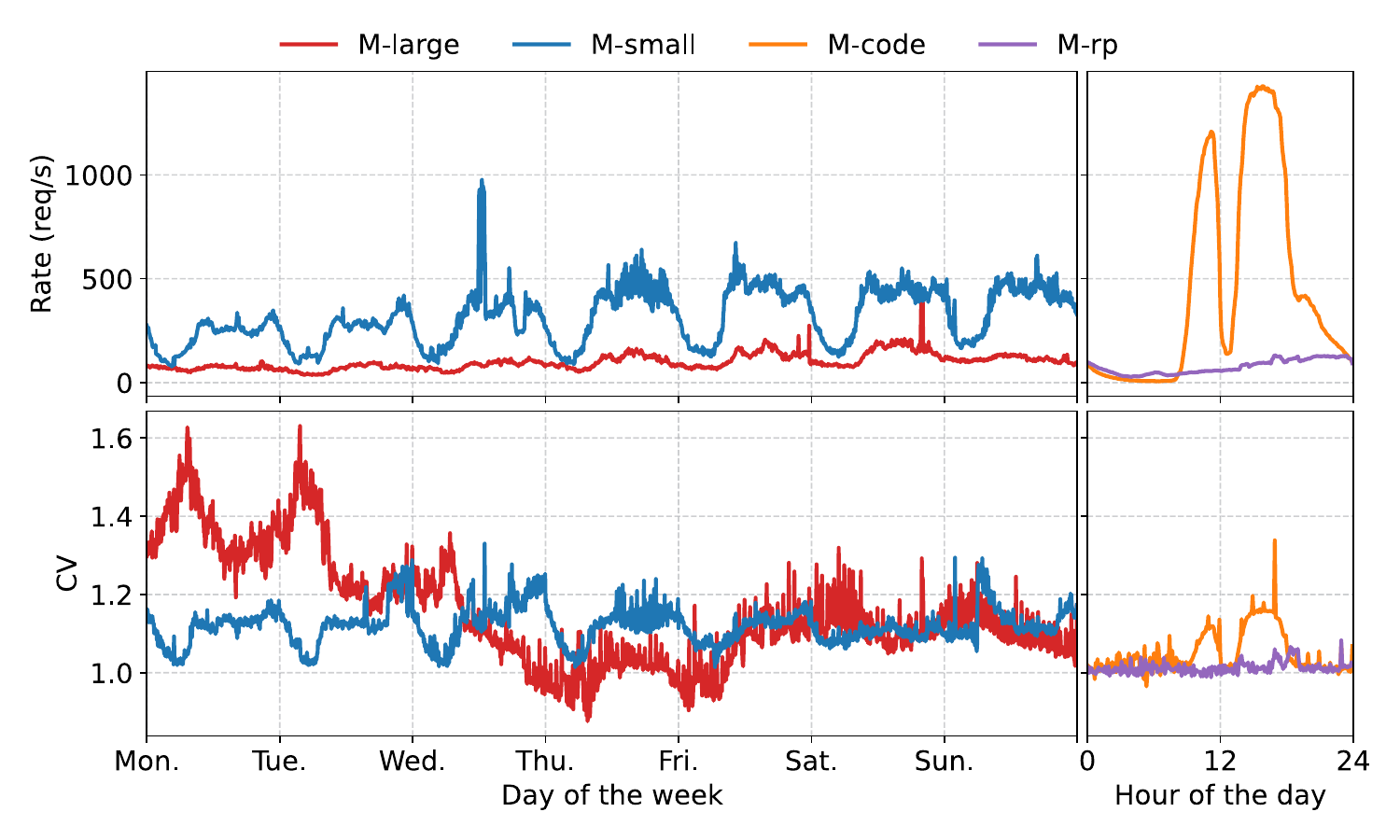}
    \vspace{-0.2in}
    \caption{
        Long-term rate and CV shifts.
    }
    \label{fig:rate-cv}
    \vspace{-0.225in}
\end{figure}

{
\begin{mdframed}[backgroundcolor=gray!10,linecolor=black,]
    \textbf{Finding~\finding{f:iat}}: The short-term arrival of LLM requests is often bursty 
    (CV $>$ 1), exhibiting complex patterns beyond any single stochastic process.
\end{mdframed}
}

\stitle{Shifting rate and burstiness.} 
Figure~\ref{fig:rate-cv} depicts the request rate and CV computed in 5-minute windows 
for multiple workloads, ranging from general-purpose (over a week) to task-specific 
ones (over a day).
We observe evident diurnal fluctuations for the arrival rate: the load peaks during 
the afternoons while dropping significantly in the early mornings, resulting in potentially 
extreme rate shifts (as shown for \lingma). 
Moreover, Figure~\ref{fig:rate-cv} also displays diverse and shifting CV patterns for 
different workloads, underscoring the instability of burstiness~\cite{burstgpt} 
in real-world workloads. 
For instance, \qmax was continuously bursty for two days (Mon. and Tue.) before turning stable (Thu. and Fri.). 
Meanwhile, request arrivals in \spark remain non-bursty for the entire day of analysis. 
We believe such diversity is partly caused by the invocation pattern 
associated with each workload: while role-playing (\spark) typically involves human 
interaction (\ie invoked via chatbots), where bursts are less common, 
general-purpose workloads (\qmax) likely include API invocations with bursts 
of batched request submission.

These shifts in rate and burstiness have strong implications for LLM 
serving system design. For one, rate shifts demonstrate the importance of \emph{auto-scaling} 
mechanisms in order to properly provision resources. 
For another, CV shifts provide both challenges and opportunities for 
designing \emph{request scheduling} policies, which should acknowledge and 
adapt to different levels of burstiness. In contrast, systems that assume static 
workload patterns may not perform well in practice.

{
\begin{mdframed}[backgroundcolor=gray!10,linecolor=black,]
    \textbf{Finding~\finding{f:ratecv}}: The arrival of LLM serving requests 
    shows a diverse shifting pattern in terms of rate and burstiness, calling 
    for adaptive system design. 
\end{mdframed}
}

\begin{figure*}[t!]
    \centering
    \begin{minipage}{0.644\textwidth}
        
        \centering
        \begin{subfigure}[t]{1\textwidth}
            \centering
            \includegraphics[width=\linewidth]{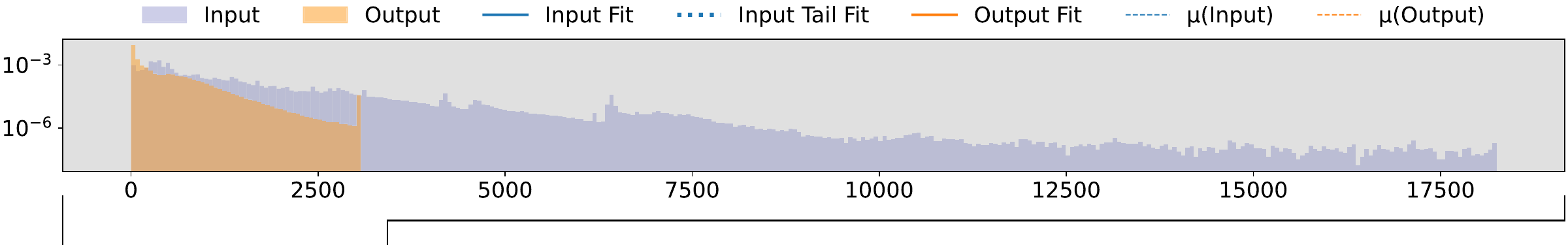}
        \end{subfigure}
        \vspace{-0.15in}

        \begin{subfigure}[t]{0.25\textwidth}
            \centering
            \includegraphics[width=\linewidth]{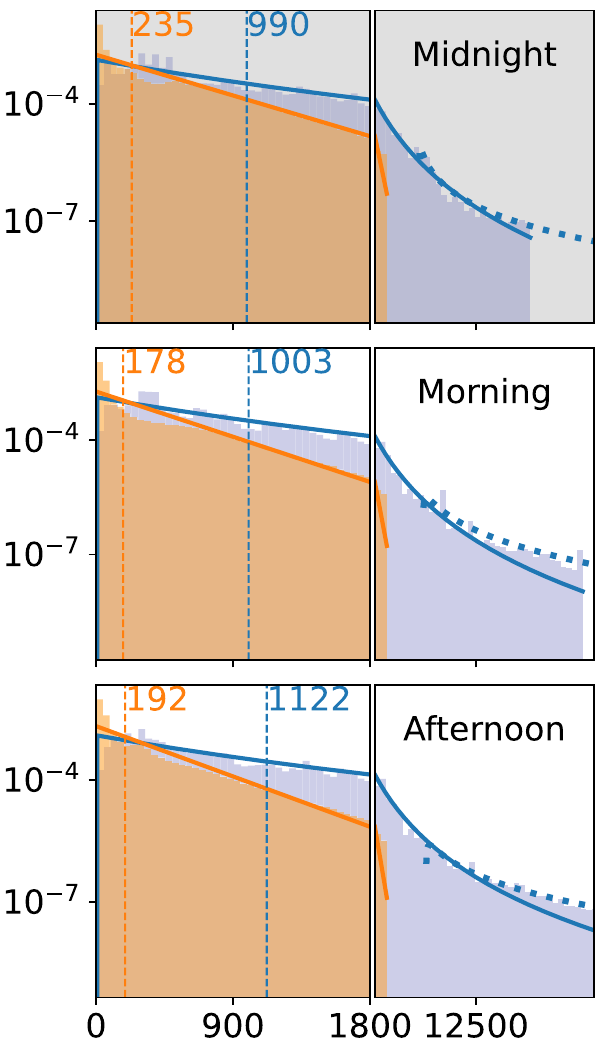}
            \vspace{-0.25in}
            \caption{
                \qplus
            }
            \label{fig:io:1}
        \end{subfigure}
        \hspace{-5pt}
        \begin{subfigure}[t]{0.25\textwidth}
            \centering
            \includegraphics[width=\linewidth]{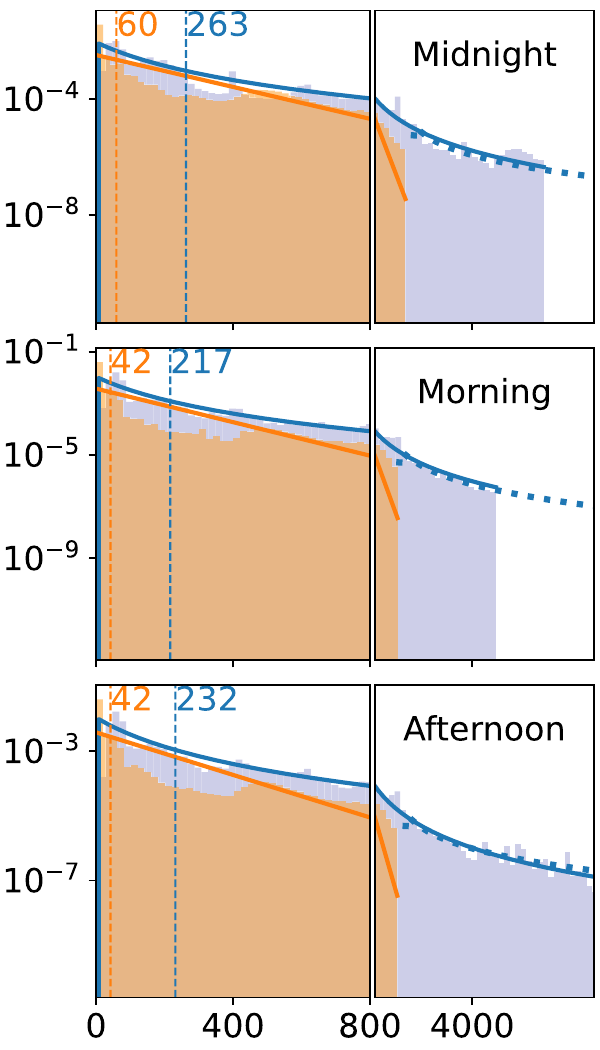}
            \vspace{-0.25in}
            \caption{
                \qturbo
            }
            \label{fig:io:2}
        \end{subfigure}
        \hspace{-5pt}
        \begin{subfigure}[t]{0.25\textwidth}
            \centering
            \includegraphics[width=\linewidth]{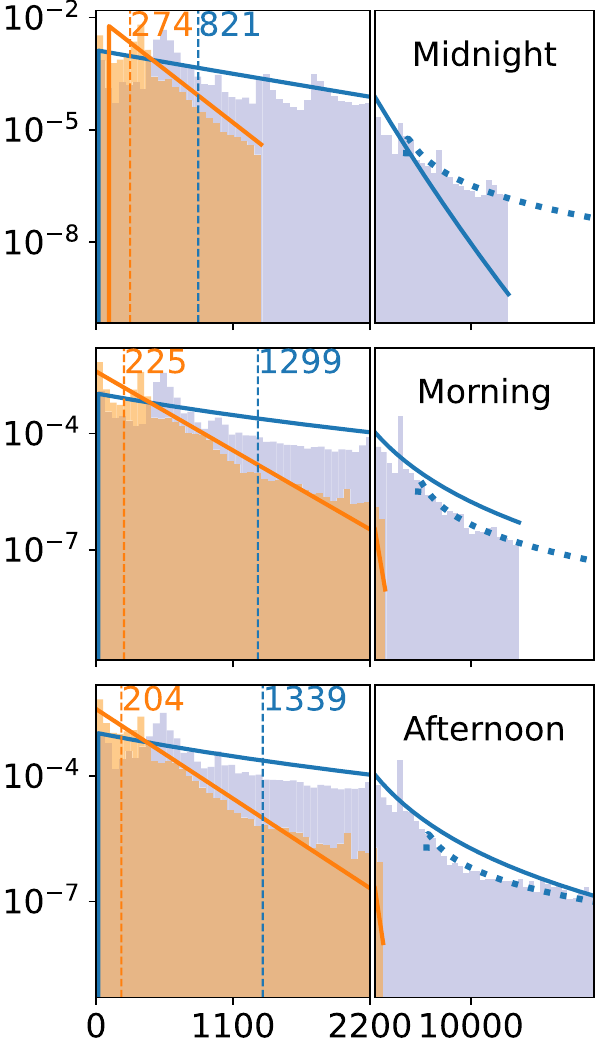}
            \vspace{-0.25in}
            \caption{
                \qlong
            }
            \label{fig:io:3}
        \end{subfigure}
        \hspace{-5pt}
        \begin{subfigure}[t]{0.25\textwidth}
            \centering
            \includegraphics[width=\linewidth]{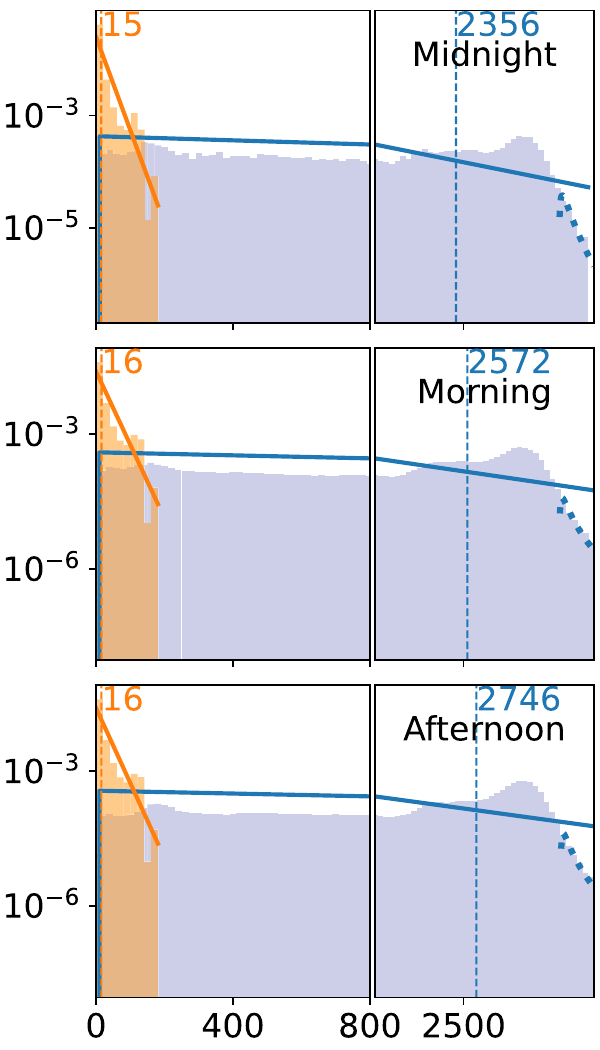}
            \vspace{-0.25in}
            \caption{
                \lingma
            }
            \label{fig:io:4}
        \end{subfigure}
        \vspace{-0.15in}
        \caption{
            Input and output length distribution. 
            \emph{x-axis}: \# tokens; \emph{y-axis}: frequency. 
            Each subfigure corresponds to a specific workload and time period, 
            split to two consecutive x-scales to show both the shift in average lengths (left) 
            as well as the tail distribution (right).
        }
        \label{fig:io}
    \end{minipage}%
    \hspace{25pt}
    \begin{minipage}{0.249\textwidth}

        \centering
        \begin{subfigure}[t]{1\textwidth}
            \centering
            \includegraphics[width=\linewidth]{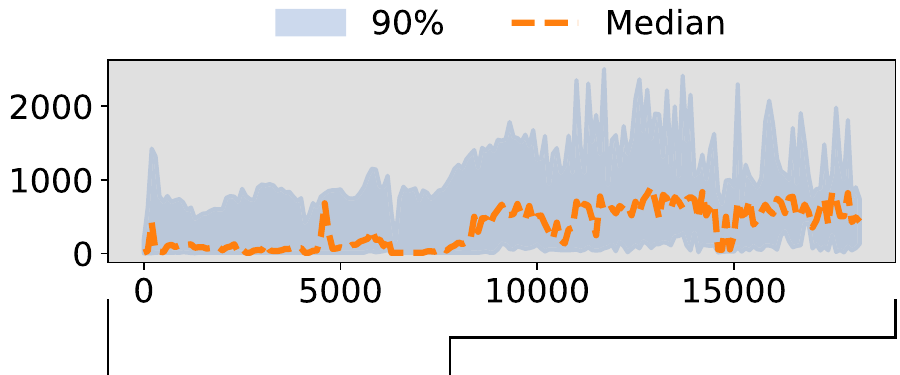}
        \end{subfigure}
        \vspace{-0.15in}

        \begin{subfigure}[t]{0.5\textwidth}
            \centering
            \includegraphics[width=\linewidth]{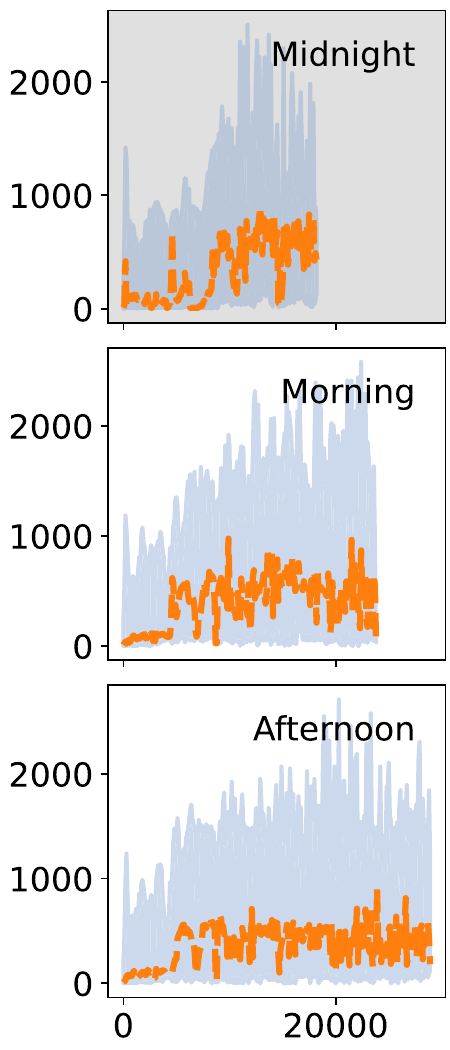}
            \label{fig:cor:1}
            \vspace{-0.25in}
            \caption{
                \qplus
            }
        \end{subfigure}
        \hspace{-5pt}
        \begin{subfigure}[t]{0.5\textwidth}
            \centering
            \includegraphics[width=\linewidth]{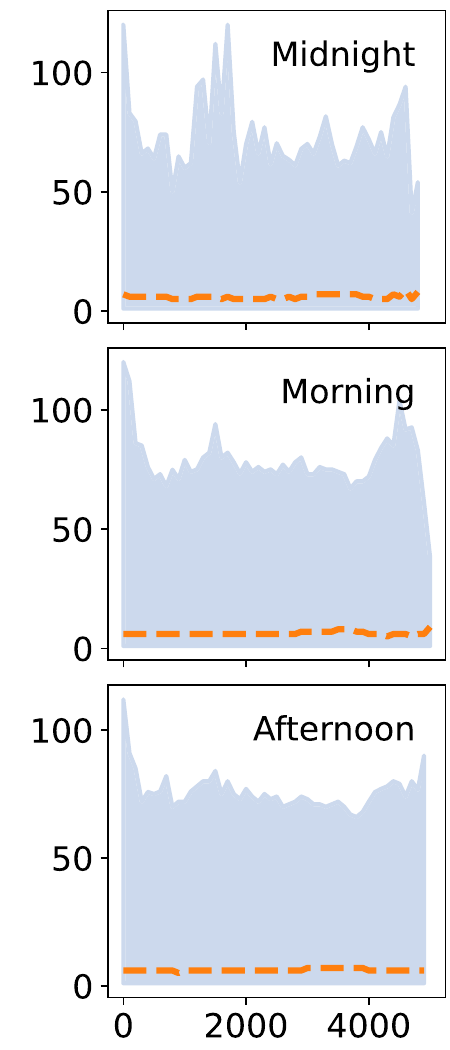}
            \label{fig:cor:2}
            \vspace{-0.25in}
            \caption{
                \lingma
            }
        \end{subfigure}
        \vspace{-0.15in}
        \caption{
            Input and output length correlation.
            \emph{x-axis}: input length; \emph{y-axis}: output length. 
        }
        \label{fig:cor}
    \end{minipage}%
    \vspace{-0.225in}
\end{figure*}

\subsection{Input and Output Length Distribution}
\label{sec:text:io}

We now characterize the input and output lengths of requests by examining their 
distributions in Figure~\ref{fig:io}. 
Figure~\ref{fig:cor} further presents the correlation between 
input and output lengths by binning similar input lengths and showing the 
90\% percentile range and median of the respective output lengths.

\stitle{Modeling length distributions.}
Existing studies~\cite{burstgpt} have advocated modeling request input lengths 
with the Zipf distribution, acknowledging an implicit power law: input lengths 
have large variance and a long upper tail (\ie requests with 
exceedingly long prompts). In our analysis, we find that input lengths in  
general-purpose workloads are best modeled by \emph{Pareto} distributions mixed 
with \emph{Log-normal} distributions (both are also power-law distributions) 
for handling the fat tail, as shown in Figure~\ref{fig:io:1} and~\ref{fig:io:2} 
with the \emph{Input Fit} and \emph{Input Tail Fit} curves. 
For task-specific workloads, the aforementioned model is less accurate due to domain-specific 
bias, such as the usage of common system prompts or templates. 

Surprisingly, we find that \emph{Exponential} distributions fit remarkably well for 
output lengths, with the only obvious exception of \qturbo in Figure~\ref{fig:io:2}. 
While it is difficult to pinpoint the exact reason behind this phenomenon 
(likely a combined result of training and workload semantics), the 
implication is worth noting: the remaining output length of an LLM request is 
not conditioned on the generated length so far, \ie the output length distribution 
is \emph{memoryless}. 

Lastly, while Figure~\ref{fig:cor} exhibits a rough positive correlation between 
input and output lengths (\ie long prompts lead to long responses), 
the relation is not as pronounced as reported in previous studies~\cite{burstgpt}. 
We believe that in practice, the correlation is diminished by complicated workload 
semantics, such as prompt templates or structured outputs.

{
\begin{mdframed}[backgroundcolor=gray!10,linecolor=black,]
    \textbf{Finding~\finding{f:io}}: The input length distribution can be modeled with  
    a mixture of Pareto and Log-normal distributions, and the output with  
    Exponential distributions. Correlation between input and output lengths is 
    weak.
\end{mdframed}
}

\stitle{Shifting length distributions.}
Motivated by Finding~\ref{f:ratecv}, we repeat the preceding analysis using data 
sampled from three different periods in a day, as shown in Figure~\ref{fig:io} and~\ref{fig:cor}. 
While the correlation appears independent of time, the actual distribution, 
contrary to common beliefs, \emph{does} shift with time. 
Notably, the range of such shifts can be up to $1.63\times$ for input (Figure~\ref{fig:io:3})  
and $1.46\times$ for output (Figure~\ref{fig:io:4}), measured by the maximal 
average length over the minimal. 

Further, input and output length shifts occur 
\emph{independently}, as demonstrated by \qplus in Figure~\ref{fig:io:1}: 
from \emph{Midnight} to \emph{Afternoon}, \qplus's input length increases 
by 13\% on average, while its output length drops by 18\%.
Intertwined with the request rate shifts, this observation translates to diverse
load on the prefill and decoding phases of LLM serving, thus directly impacting system performance. 
Non-disaggregated serving systems may face variable performance interference 
between the two phases~\cite{distserve}, while disaggregated systems might require 
phase-independent resource auto-scaling.

{
\begin{mdframed}[backgroundcolor=gray!10,linecolor=black,]
    \textbf{Finding~\finding{f:ioshift}}: The input and output length distributions shift 
    dynamically and independently over time, leading to diverse load fluctuations 
    for prefill and decoding.
\end{mdframed}
}

\subsection{Client Decomposition}
\label{sec:text:client}

\begin{figure*}[t!]
    \centering
    \begin{minipage}{0.252\textwidth}
        
        \begin{subfigure}[t]{1\textwidth}
            \centering
            \includegraphics[width=\linewidth]{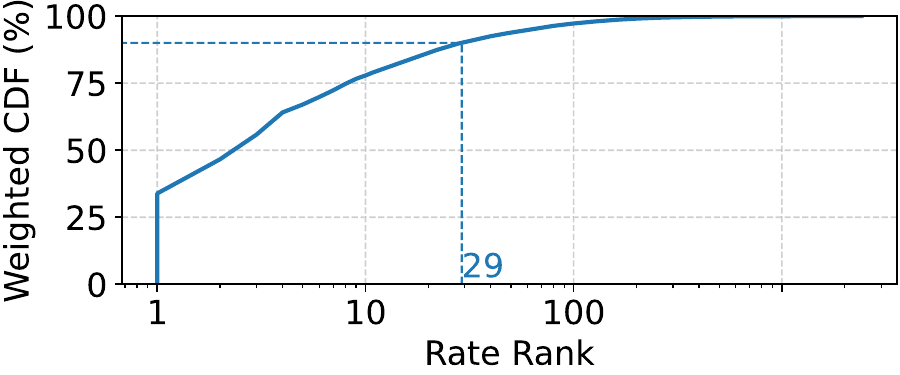}
            \vspace{-0.25in}
        \end{subfigure}
        \begin{subfigure}[t]{1\textwidth}
            \centering
            \includegraphics[width=\linewidth]{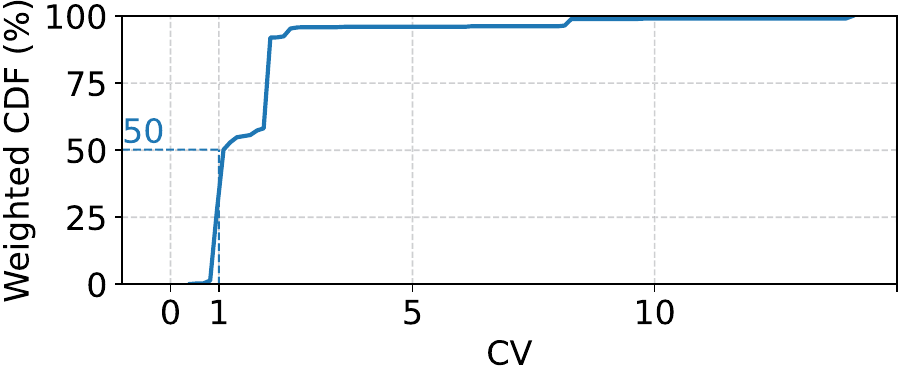}
            \vspace{-0.25in}
        \end{subfigure}
        \begin{subfigure}[t]{1\textwidth}
            \centering
            \includegraphics[width=\linewidth]{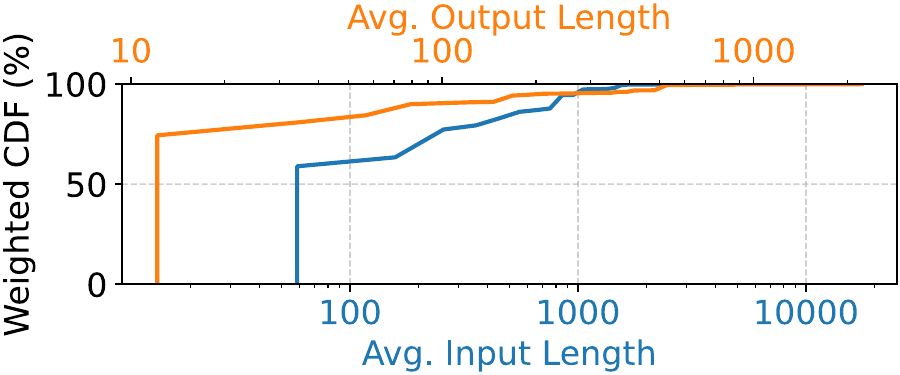}
            \vspace{-0.25in}
        \end{subfigure}
        \vspace{-0.3in}
        \caption{
            Client heterogeneity in terms of rate, \etc
            All CDFs are weighted by client rates.  
        }
        \label{fig:hetero}
    \end{minipage}%
    \hspace{20pt}
    \begin{minipage}{0.652\textwidth}
        \begin{subfigure}[t]{1\textwidth}
            \centering
            \includegraphics[width=\linewidth]{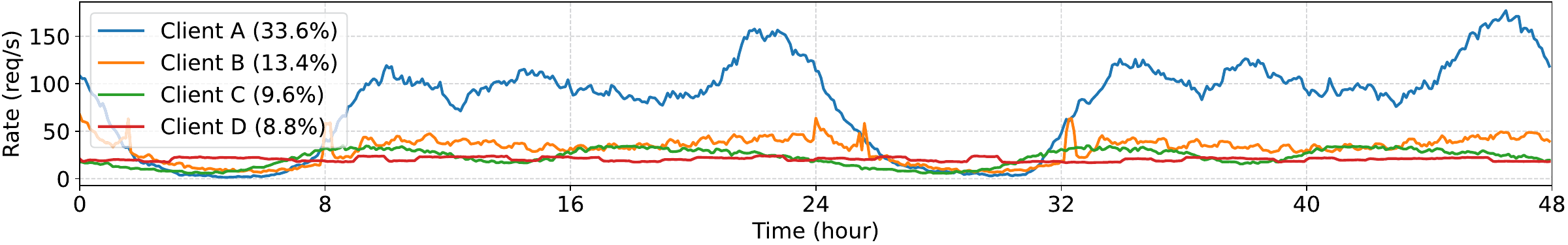}
            \vspace{-0.25in}
        \end{subfigure}
        \begin{subfigure}[t]{1\textwidth}
            \centering
            \includegraphics[width=\linewidth]{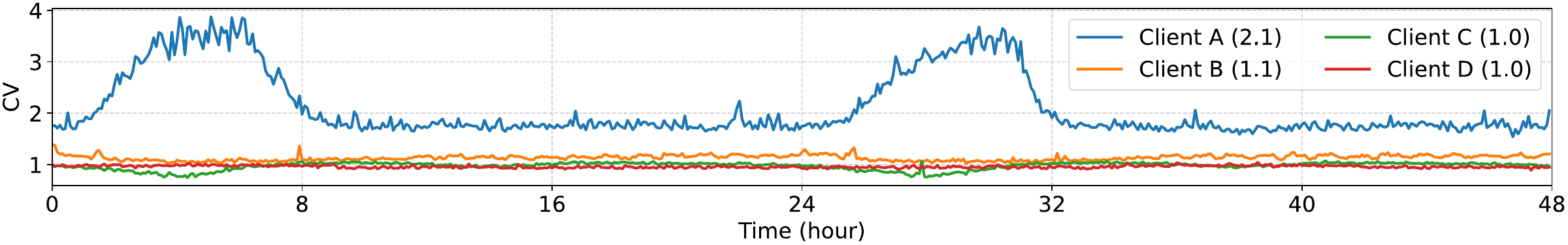}
            \vspace{-0.25in}
        \end{subfigure}

        \begin{subfigure}[t]{0.49375\textwidth}
            \centering
            \includegraphics[width=\linewidth]{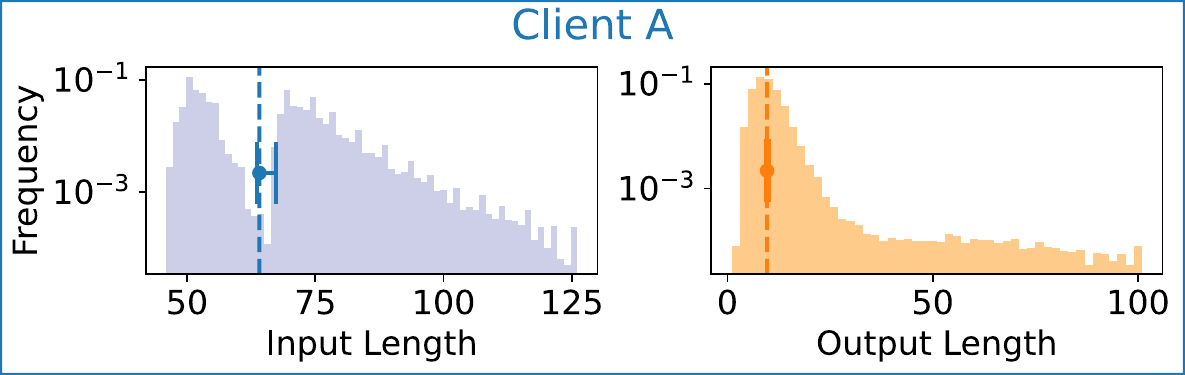}
            \vspace{-0.25in}
        \end{subfigure}
        \hfill
        \begin{subfigure}[t]{0.49375\textwidth}
            \centering
            \includegraphics[width=\linewidth]{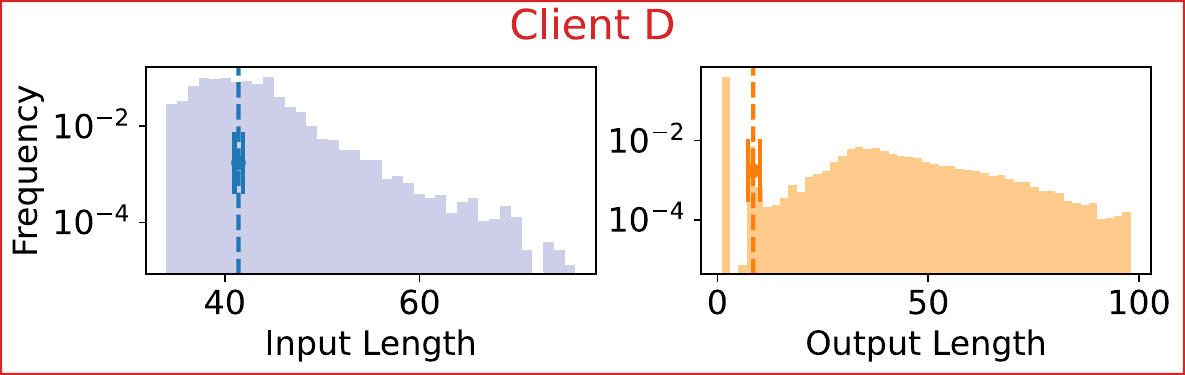}
            \vspace{-0.25in}
        \end{subfigure}
        \vspace{-0.25in}
        \caption{
            Characterization of the top four clients in the \qturbo workload in isolation, 
            using the first 48-hour data from Figure~\ref{fig:rate-cv}.  
            Vertical lines in the last-row subfigures indicate average input/output 
            lengths, and error bars show the range of average lengths in 1-hour windows.
        }
        \label{fig:top}
    \end{minipage}%
    \vspace{-0.175in}
\end{figure*}

Thus far, we have uncovered several shifting patterns in LLM serving 
workloads with concrete real-world implications. 
These patterns are non-trivial to model because they are the aggregate behavior of 
multiple \emph{clients}, each corresponding to an end user or 
upstream application (\eg a chatbot that relies on our service). 
For more insights, we conduct a decomposition analysis of 
the \qturbo workload on a \emph{per-client} basis.
We report substantial heterogeneity and stability in client behaviors, and further 
reveal that the shifting patterns are largely attributable to rate 
fluctuations among top clients. 

\stitle{Client heterogeneity and stability.} 
Figure~\ref{fig:hetero} characterizes client behaviors in terms of their 
rate, burstiness, and request length distribution, using the first 48-hour 
data from Figure~\ref{fig:rate-cv}.
We observe highly skewed client rates: out of 2,412 clients, the top 29 clients (ranked 
by their rate in descending order) are responsible for 90\% of the requests. 
Furthermore, client burstiness and input/output lengths span a 
diverse range, indicating the fundamental heterogeneity of clients. 

Meanwhile, in isolation, top clients exhibit notable stability in all 
aspects other than their request rate, as shown in Figure~\ref{fig:top}.
For example, the burstiness of Clients B, C, and D remains mostly stable within 48 
hours, and the burstiness of Client A only deviates in the early mornings when the  
rate drops exceedingly low. Additionally, Clients A and D display stable input and 
output lengths, as indicated by the small error bars in the last-row subfigures, 
which visualize the range of average lengths during the entire period of our analysis.

\stitle{Impact of top clients.} 
These observations suggests the following \emph{causal modeling}: 
characteristics of the whole workload are steered by a few top clients, 
whose rate fluctuations effectively cause the workload to shift towards different patterns. 

This modeling indeed accounts for many previously found patterns in \qturbo. 
For example, note that in Figure~\ref{fig:rate-cv}, the workload temporarily 
bursts on Tuesday night. This matches the fact that in Figure~\ref{fig:top}, 
the rate of Client A (which is bursty) also sees a peak at around the same time. 
As for request lengths, the average input length of \qturbo decreases from 
\emph{Midnight} to \emph{Morning} in Figure~\ref{fig:io}, 
aligning with the increase of request rate from Client A (whose input lengths are 
shorter than average) from hour 1 to hour 9 in Figure~\ref{fig:top}. 
We rely on the same causal modeling to generate realistic workloads 
that encompass the intricate shifting patterns in \S\ref{sec:reconstruct}, 
where we evaluate the accuracy and benefits of our approach quantitatively. 

{
\smallskip
\begin{mdframed}[backgroundcolor=gray!10,linecolor=black,]
    \textbf{Finding~\finding{f:client}}: Real-world workloads consist of heterogeneous 
    clients with skewed arrival rates. The top clients
    and their rate fluctuations largely explain the shifting workload patterns. 
\end{mdframed}
}
\section{Characterizing Multimodal Workloads}
\label{sec:mm}

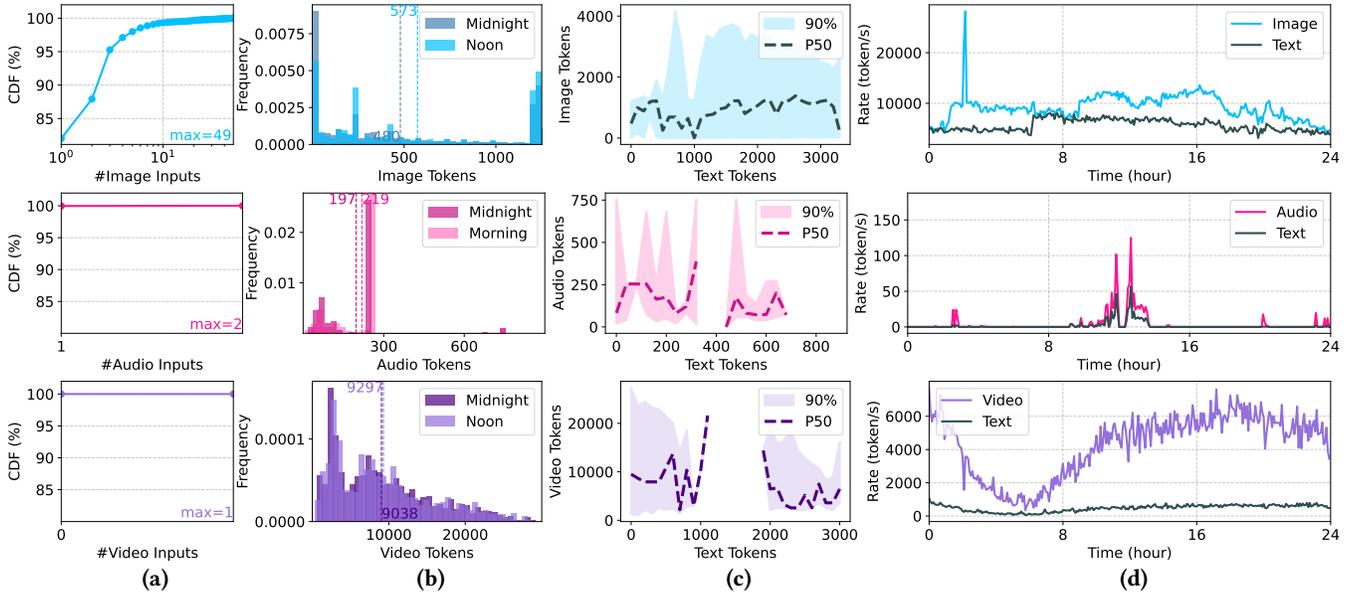
\begin{figure*}[t!]
    \centering
    \begin{subfigure}[t!]{.94\textwidth}
      \centering
      \includegraphics[width=\linewidth]{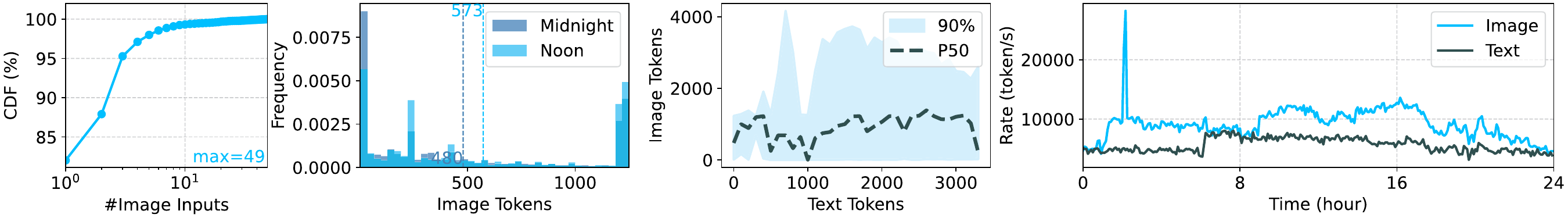}
    \end{subfigure}
    \begin{subfigure}[t!]{.94\textwidth}
        \centering
        \includegraphics[width=\linewidth]{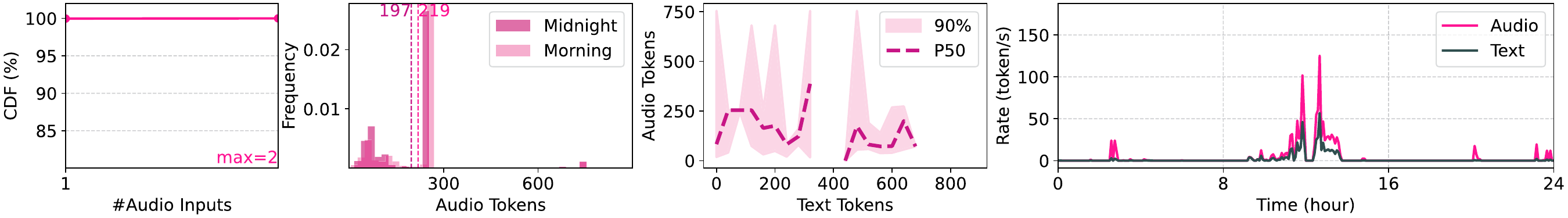}
    \end{subfigure}
    \begin{subfigure}[t!]{.94\textwidth}
        \centering
        \includegraphics[width=\linewidth]{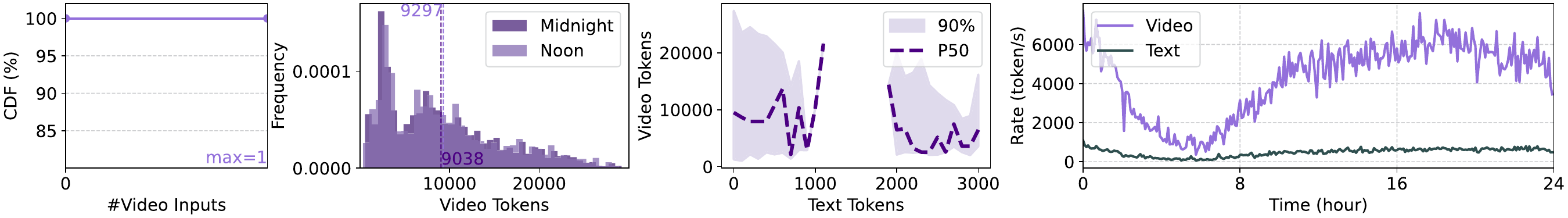}
    \end{subfigure}
    \begin{picture}(500,6)
       \small
        \put(50,-1.3){\textbf{(a)}}
        \put(154,-1.3){\textbf{(b)}}
        \put(271,-1.3){\textbf{(c)}}
        \put(420,-1.3){\textbf{(d)}}
    \end{picture}
    \vspace{-0.1in}
    \caption{
        Characterization of multimodal inputs in \image, \audio, and \video (\emph{Rows}). 
        \emph{Columns}: 
        (a) \#multimodal inputs per request; 
        (b) length distribution of inputs;
        (c) correlation between text and multimodal tokens;
        (d) overall arrival rate of multimodal and text tokens. 
    }
    \label{fig:mm:data}
    \vspace{-0.175in}
\end{figure*}

We next examine workloads for multimodal models.
Our characterization reveals that the tokenized length distributions are 
irregular across image, audio, and video modalities, contributing to highly variable 
modal load that also shifts over time (\S\ref{sec:mm:data}). 
Together with overhead from downloading, normalizing, and encoding (\S\ref{sec:mm:hetero}), 
multimodal inference is prone to considerable request \emph{heterogeneity} between modalities, 
leading to prolonged time-to-first-token (TTFT).
We report how client decomposition helps capture these patterns 
and facilitates a deeper understanding of multimodal workloads (\S\ref{sec:mm:client}).

\subsection{Modality Load Variance}
\label{sec:mm:data}

\stitle{Load variance in different modalities.}
Figure~\ref{fig:mm:data} characterizes data distributions in \image, \audio, and \video,
focusing specifically on the multimodal inputs. 
Unlike text prompts, multimodal inputs are more likely to have standard sizes 
depending on upstream applications. 
As such, in all three workloads, the tokenized lengths of multimodal inputs exhibit 
irregularly shaped distributions, clustering around certain values (\eg around 2,500 
for \video in (b)) instead of following typical power-law distributions like the
text modality (see Figure~\ref{fig:io}).
In addition, given the diverse number of multimodal inputs per request (shown in (a)) 
and the lack of correlation between text and multimodal tokens (shown in (c)), 
we observe highly \emph{varied} load on modality encoders, as illustrated in (d).

\begin{figure}[t!]
    \centering
    \includegraphics[width=.95\linewidth]{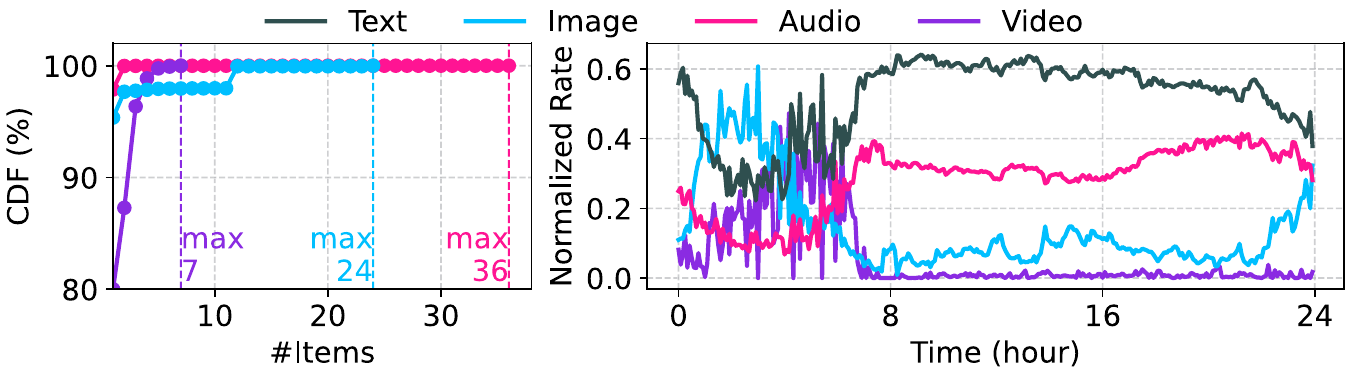}
    \vspace{-0.175in}
    \caption{
        Characterization of omni-modal inputs in \omni.
        \emph{Left}: number of multimodal inputs per request.
        \emph{Right}: arrival rate of multimodal and text tokens, normalized 
        by the total input rate.
    }
    \label{fig:mm:omni}
    \vspace{-0.15in}
\end{figure}

\begin{figure}[t!]
    \begin{subfigure}[t!]{.315\linewidth}
        \centering
        \includegraphics[width=\linewidth]{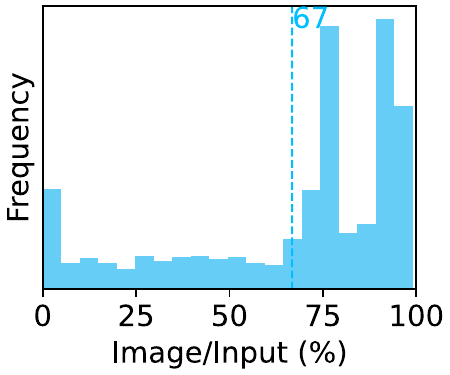}
    \end{subfigure}
    \begin{subfigure}[t!]{.315\linewidth}
        \centering
        \includegraphics[width=\linewidth]{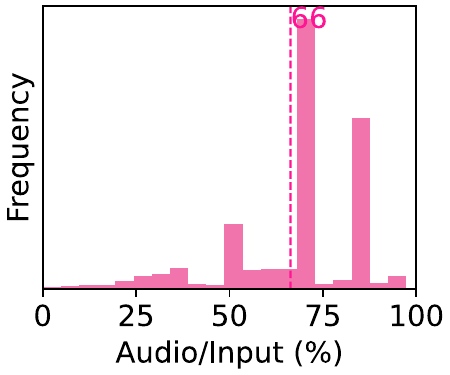}
    \end{subfigure}
    \begin{subfigure}[t!]{.315\linewidth}
        \centering
        \includegraphics[width=\linewidth]{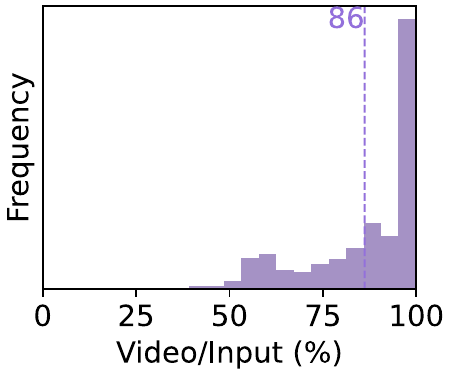}
    \end{subfigure}
    \vspace{-0.1in}
    \caption{
        Ratio of multimodal input tokens per request in \image, \audio, and \video. 
        Numbers indicate the average ratio.
    }
    \label{fig:mm:breakdown}
    \vspace{-0.2in}
\end{figure}

Two observations further complicate the load variance in multimodal workloads.  
$(i)$ The variance of multimodal load can be independent of the textual load.
For example, nine hours into the \image workload, the image token rate sees 
an abrupt increase, while the text token rate remains constant.
$(ii)$ Similar to the input/output distributions in language workloads, the multimodal 
data distributions also shift over time, as revealed in Figure~\ref{fig:mm:data}(b).  
For instance, the average image length in \image varies by up to 19\% over one day.

\stitle{Load variance in omni-modality.}
Figure~\ref{fig:mm:omni} presents the same analysis on \omni, an omni-modal 
workload where requests can contain more than two modalities.
Unsurprisingly, the workload exhibits more complex variability, featuring a greater number 
of multimodal inputs per request and more diverse shifting patterns in input load
(\eg audio load rises during the day, while image load becomes prominent past midnight). 
Moreover, as new applications of omni-modal LLMs emerge, 
we anticipate that the load variance in omni-modal workloads will continue to evolve.

In both cases, load variance presents challenges to the resource efficiency 
of multimodal inference, necessitating serving systems with more flexible resource scaling.

{
\smallskip
\begin{mdframed}[backgroundcolor=gray!10,linecolor=black,]
    \textbf{Finding~\finding{f:mm-data}}: Multimodal data distributions exhibit 
    irregular and independent shifts, underscoring significant load variance 
    across modalities. 
\end{mdframed}
}

\subsection{Request Heterogeneity}
\label{sec:mm:hetero}

Multimodal inputs introduce complexity not only to the overall load, but also to individual 
requests.
Figure~\ref{fig:mm:breakdown} presents a breakdown of each request's input tokens in the 
\image, \audio, and \video workloads, revealing a flat distribution in every case. This indicates that 
real multimodal requests are \emph{heterogeneous}, naturally ranging from text-heavy to 
multimodal-heavy in terms of input composition.

\begin{figure*}[t!]
    \centering
    \begin{subfigure}[t!]{.45\textwidth}
      \centering
      \includegraphics[width=\textwidth]{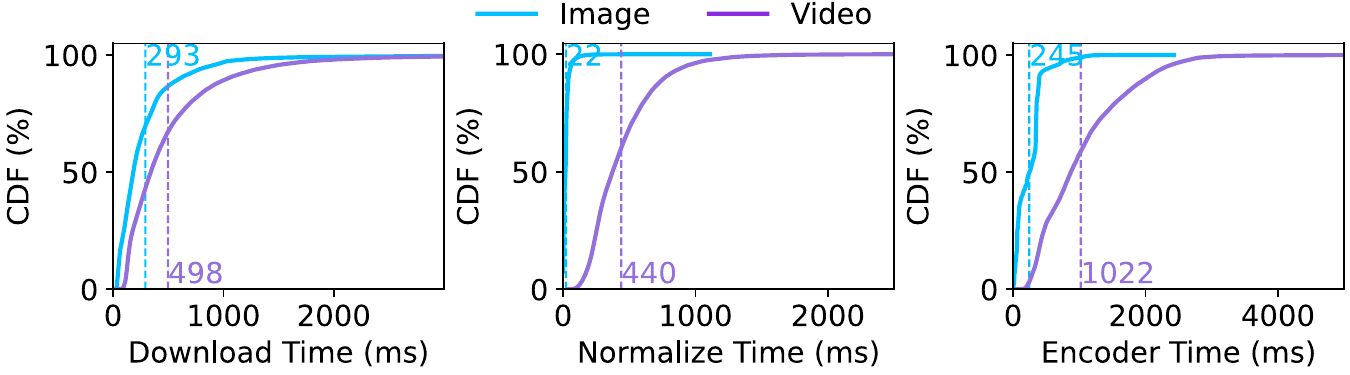}
      \caption{Per-stage time during first-token generation.}
      \label{fig:mm:time:stage}
    \end{subfigure}
    \begin{subfigure}[t!]{.45\textwidth}
        \centering
        \includegraphics[width=\textwidth]{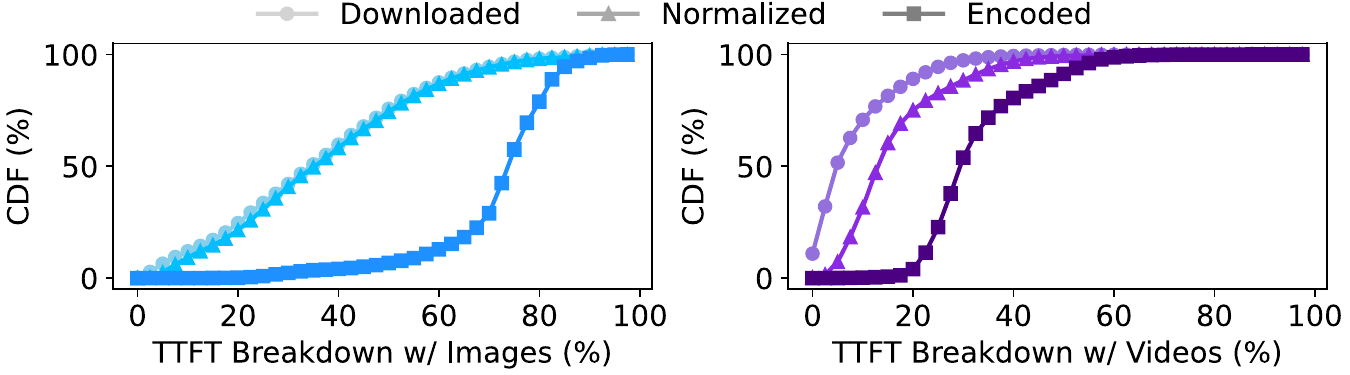}
        \caption{CDF of cummulative time after each stage.}
      \label{fig:mm:time:ttft}
    \end{subfigure}
    \vspace{-0.05in}
    \caption{
        Breakdown of first-token time when serving requests 
        with image or video inputs (\image and \video).
    }
    \label{fig:mm:time}
    \vspace{-0.1in}
\end{figure*}

In practice, such heterogeneity is challenging for serving systems, as it translates 
into prolonged TTFT during inference, as shown in Figure~\ref{fig:mm:time}. 
On one hand, for multimodal-heavy requests, the \emph{download}, \emph{normalization}, 
and \emph{encoding} stages for tokenizing multimodal inputs all contribute to 
considerable extra overhead (reported in Figure~\ref{fig:mm:time:stage}) 
in the first-token generation process, directly lengthening the TTFT, as illustrated in 
Figure~\ref{fig:mm:time:ttft}. 
For instance, half of the \image requests spend 75\% of their TTFT before LLM prefilling.
On the other hand, the extremely long-tailed distribution of encoder time 
in Figure~\ref{fig:mm:time} signifies potential queuing that affects 
text-heavy requests as well.
For example, a request with few image tokens in \image may be blocked at the 
encoding stage by previously scheduled image-heavy requests; 
or it may experience slower encoding due to suboptimal batching that only 
considers prefill execution.
This highlights the need for more advanced scheduling and batching strategies.

{
\begin{mdframed}[backgroundcolor=gray!10,linecolor=black,]
    \textbf{Finding~\finding{f:mm-breakdown}}: Multimodal requests are heterogeneous 
    with diverse ratios of multimodal inputs per request, leading to prolonged 
    TTFTs that require tailored optimizations.
\end{mdframed}
}

\subsection{Mulitmodal Client Decomposition}
\label{sec:mm:client}

Given the involved patterns in multimodal workloads, we present a 
client decomposition of \image similar to that in \S\ref{sec:text:client} to 
further complement our characterization. 
Notably, our causal modeling proposed by Finding~\ref{f:client} 
still applies, as we verify that load variance and request heterogeneity 
are explainable by the multimodal client behaviors.

\begin{figure*}[t!]
    \centering
    \begin{minipage}[t]{0.252\textwidth}
        \centering
        \begin{subfigure}[t]{0.49\textwidth}
            \centering
            \includegraphics[width=\linewidth]{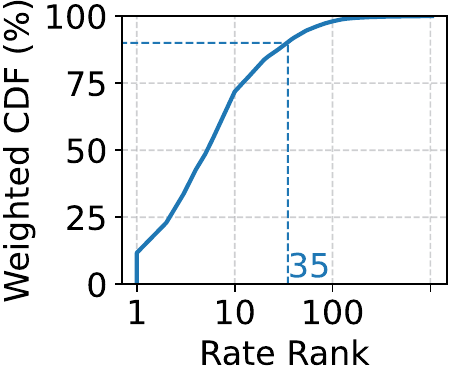}
        \end{subfigure}
        \hfill
        \begin{subfigure}[t]{0.49\textwidth}
            \centering
            \includegraphics[width=\linewidth]{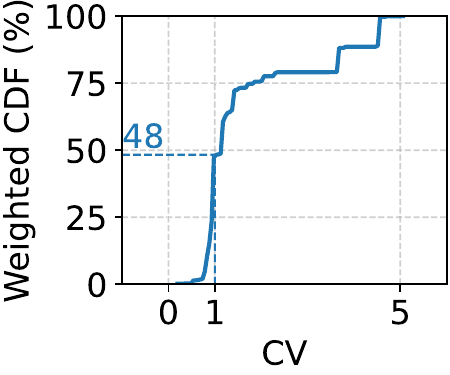}
        \end{subfigure}
        \begin{subfigure}[t]{0.49\textwidth}
            \centering
            \includegraphics[width=\linewidth]{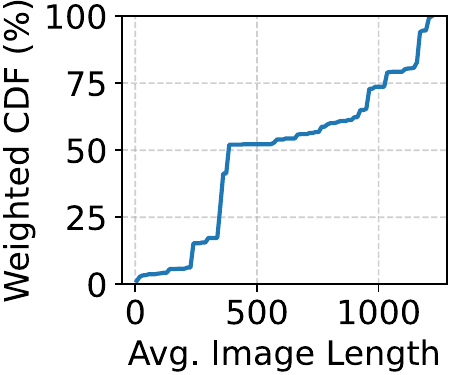}
        \end{subfigure}
        \hfill
        \begin{subfigure}[t]{0.49\textwidth}
            \centering
            \includegraphics[width=\linewidth]{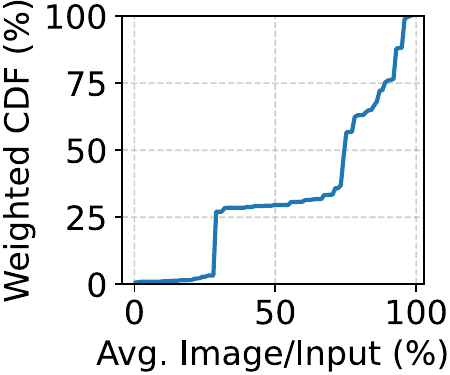}
        \end{subfigure}
        \vspace{-0.19in}
        \caption{
            Clients in \image.
            CDFs are weighted by rates.  
        }
        \label{fig:mm:client-cdf}
    \end{minipage}%
    \hspace{10pt}
    \begin{minipage}[t]{0.639\textwidth}
        \centering

        \begin{subfigure}[t]{1\textwidth}
            \centering
            \includegraphics[width=\linewidth]{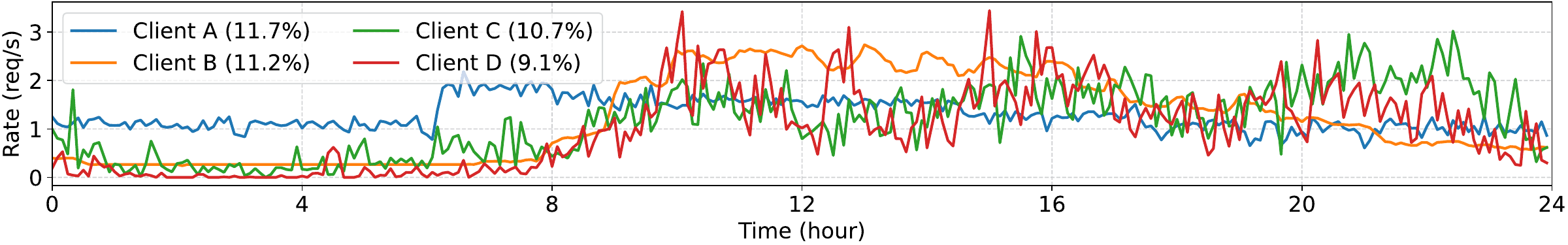}
            \vspace{-0.25in}
        \end{subfigure}
        \begin{subfigure}[t]{0.49375\textwidth}
            \centering
            \includegraphics[width=\linewidth]{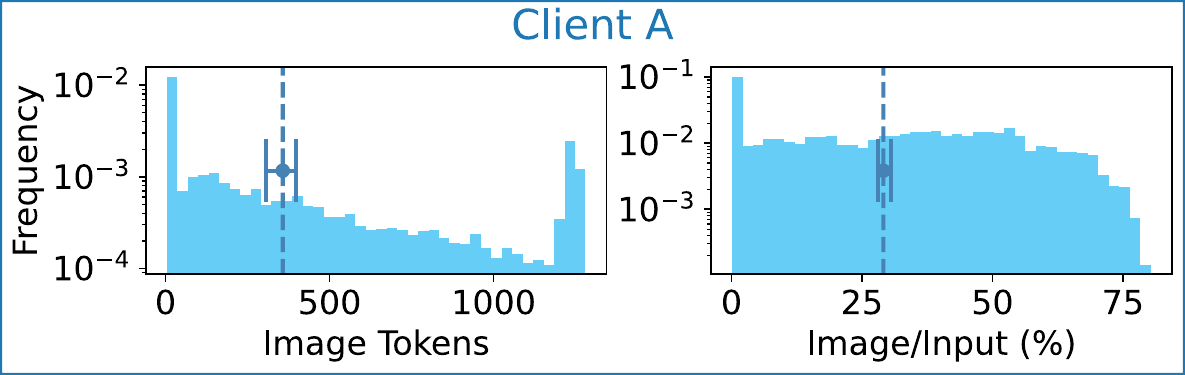}
            \vspace{-0.25in}
        \end{subfigure}
        \hfill
        \begin{subfigure}[t]{0.49375\textwidth}
            \centering
            \includegraphics[width=\linewidth]{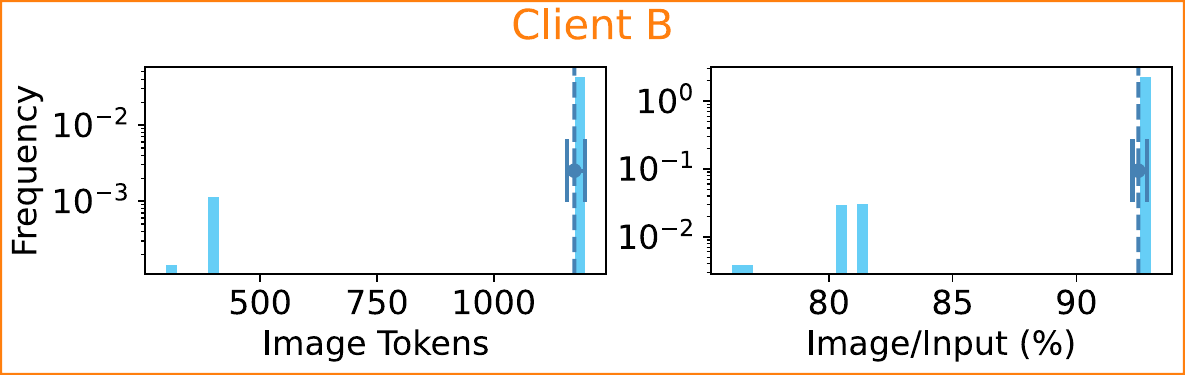}
            \vspace{-0.25in}
        \end{subfigure}
        \caption{
            Behavior of top clients in \image.
            Vertical lines in the last-row subfigures indicate average lengths, 
            and error bars show the range of average lengths within a day. 
        }
        \label{fig:mm:client-top}
    \end{minipage}%
    \vspace{-0.15in}
\end{figure*}

\stitle{Characterization of multimodal clients.}
Figure~\ref{fig:mm:client-cdf} summarizes the behaviors of 1,036 multimodal clients 
in \image, which are heterogeneous in terms of rate, burstiness, image length 
distributions, and image-to-input ratios per request.
Interestingly, the last two CDFs concerning image data in Figure~\ref{fig:mm:client-cdf} 
exhibit a \emph{staircase-like} pattern, hinting at the existence of text-heavy or multimodal-heavy clients.

Indeed, some of the top clients show remarkably skewed data distributions, 
as represented by Client B in Figure~\ref{fig:mm:client-top}, who exclusively sends images of the same size (around 1,200 tokens each) 
and requests that are similarly structured for the entire 24 hours during our measurement.
In general, top-client behaviors remain stable and predictable, as indicated by the 
narrow error bars in the lower part of Figure~\ref{fig:mm:client-top}.

\stitle{Explaining workload patterns.}
We emphasize that the top clients presented in Figure~\ref{fig:mm:client-top} have a
direct impact on the previously presented workload patterns. 
For one, the heterogeneity of clients directly contributes to the diverse image-to-input 
ratios across all requests. 
For another, Client B's rate ramps up nine hours into the workload, 
resulting in a surge of image-heavy requests (typical for this client) that exactly 
matches the increase of image load as shown in \S\ref{sec:mm:data}. 

{
\begin{mdframed}[backgroundcolor=gray!10,linecolor=black,]
    \textbf{Finding~\finding{f:mm-client}}: Top clients in multimodal workloads 
    exhibit diverse behaviors, and characterizing them helps explain
    the overall workload patterns.
\end{mdframed}
}

\section{Characterizing Reasoning Workloads}
\label{sec:reasoning}

This section focuses on analyzing reasoning workloads.
Our characterization shows that the unique ``thinking'' behavior of reasoning models (\S\ref{sec:background:llm}) 
results in longer, more variable output lengths and a distinct ratio of 
\emph{reason} and \emph{answer} tokens (\S\ref{sec:reasoning:data}). 
In addition, request arrivals in reasoning workloads are less bursty, partly owing to 
a considerable proportion of multi-turn conversations, which alter the request 
arrival pattern (\S\ref{sec:reasoning:multi-turn}).
We conclude with client decomposition to extend our causal modeling 
to reasoning workloads (\S\ref{sec:reasoning:client}).

\subsection{Understanding Reason \& Answer Lengths}
\label{sec:reasoning:data}

\begin{figure}[t!]
    \centering
    \begin{subfigure}[t]{1\linewidth}
      \centering
      \includegraphics[width=.87\linewidth]{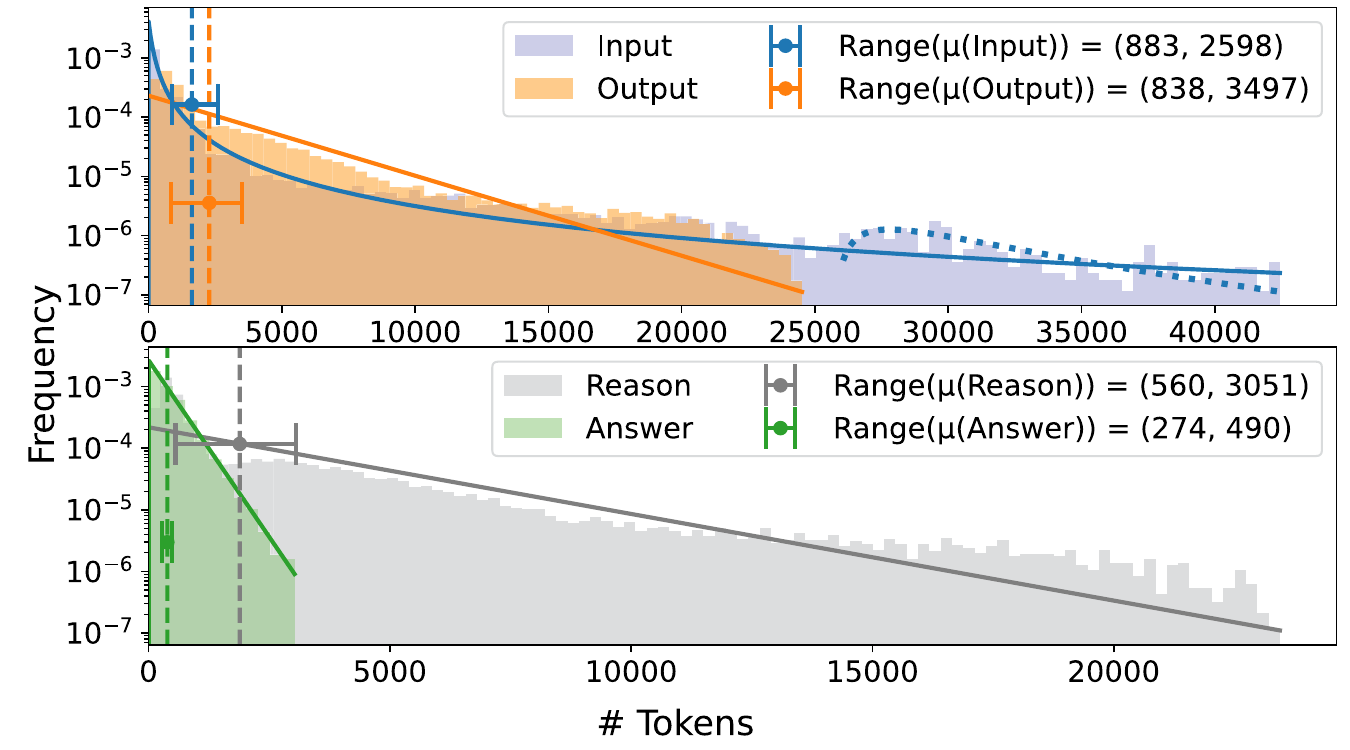}
      \vspace{-0.1in}
      \caption{Input and output length distribution, in one-hour windows.}
      \label{fig:reasoning-data:dist}
    \end{subfigure}

    \centering
    \begin{subfigure}[t]{0.44\linewidth}
        \centering
        \includegraphics[width=\linewidth]{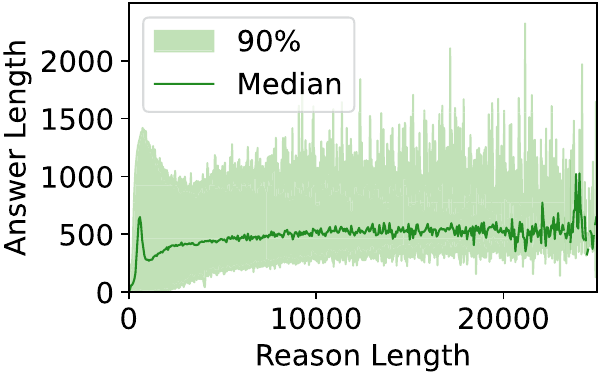}
        \vspace{-0.25in}
        \caption{Breakdown correlation.}
        \label{fig:reasoning-data:cor}
    \end{subfigure}
    \begin{subfigure}[t]{0.45\linewidth}
        \centering
        \includegraphics[width=\linewidth]{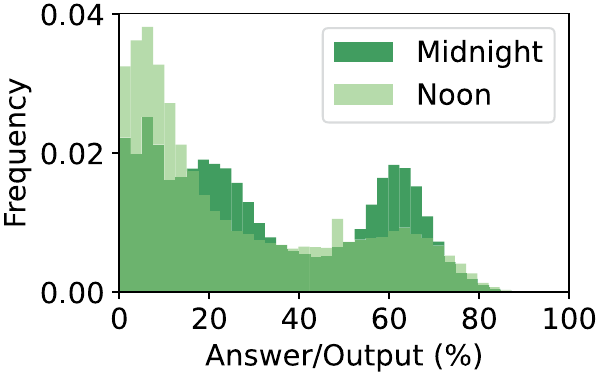}
        \vspace{-0.25in}
        \caption{Output length breakdown.}
        \label{fig:reasoning-data:ratio}
    \end{subfigure}
    \vspace{-0.05in}
    \caption{
        Characterization of input and output lengths for the \ds workload in one day.
        Error bars in (a) indicate the range of average lengths over the day. 
    }
    \label{fig:reasoning-data}
    \vspace{-0.25in}
\end{figure}

Figure~\ref{fig:reasoning-data} characterizes request lengths in the \ds workload, 
depicting also the reason and answer parts of outputs. 
In the upper part of Figure~\ref{fig:reasoning-data:dist}, we observe similar power-law 
distributions and shifting patterns (Finding \ref{f:io} and~\ref{f:ioshift}) 
in terms of input and output lengths, as indicated by the fitting curves and 
error bars.
However, output lengths are significantly longer and more variable than those 
in non-reasoning workloads, due to the long reason lengths. 
In fact, as shown in the lower part of Figure~\ref{fig:reasoning-data:dist}, 
reason lengths can be on average $4\times$ longer than answer lengths, and contribute more to the 
shifting of output lengths. The different matching levels of the Exponential 
curves suggest that, to some extent, the reasoning output behaves 
more like further input for LLMs, while the answer section remains akin to 
traditional output. 

Moreover, Figures~\ref{fig:reasoning-data:cor} and~\ref{fig:reasoning-data:ratio} 
reveal a non-trivial relation between reason and answer lengths: there exists a 
clearer correlation between them (compared with Figure~\ref{fig:cor}), 
while their per-request ratio exhibits a consistent \emph{bimodal} distribution. 
The bimodality originates from two dominating task patterns adopted by a 
reasoning model (\ie reasoning for either a more complete or more concise answer), 
which future serving optimizations may be able to leverage. 

{
\smallskip
\begin{mdframed}[backgroundcolor=gray!10,linecolor=black,]
    \textbf{Finding~\finding{f:reason-data}}: Reasoning workloads exhibit 
    longer and more variable output lengths, due to the reason tokens. 
    In relation, reason and answer lengths display stronger positive correlation, 
    as well as a unique bimodal ratio.
\end{mdframed}
}

\subsection{Arrival and Multi-Turn Patterns}
\label{sec:reasoning:multi-turn}

\begin{figure}[t!]
    \centering
    \includegraphics[width=.94\linewidth]{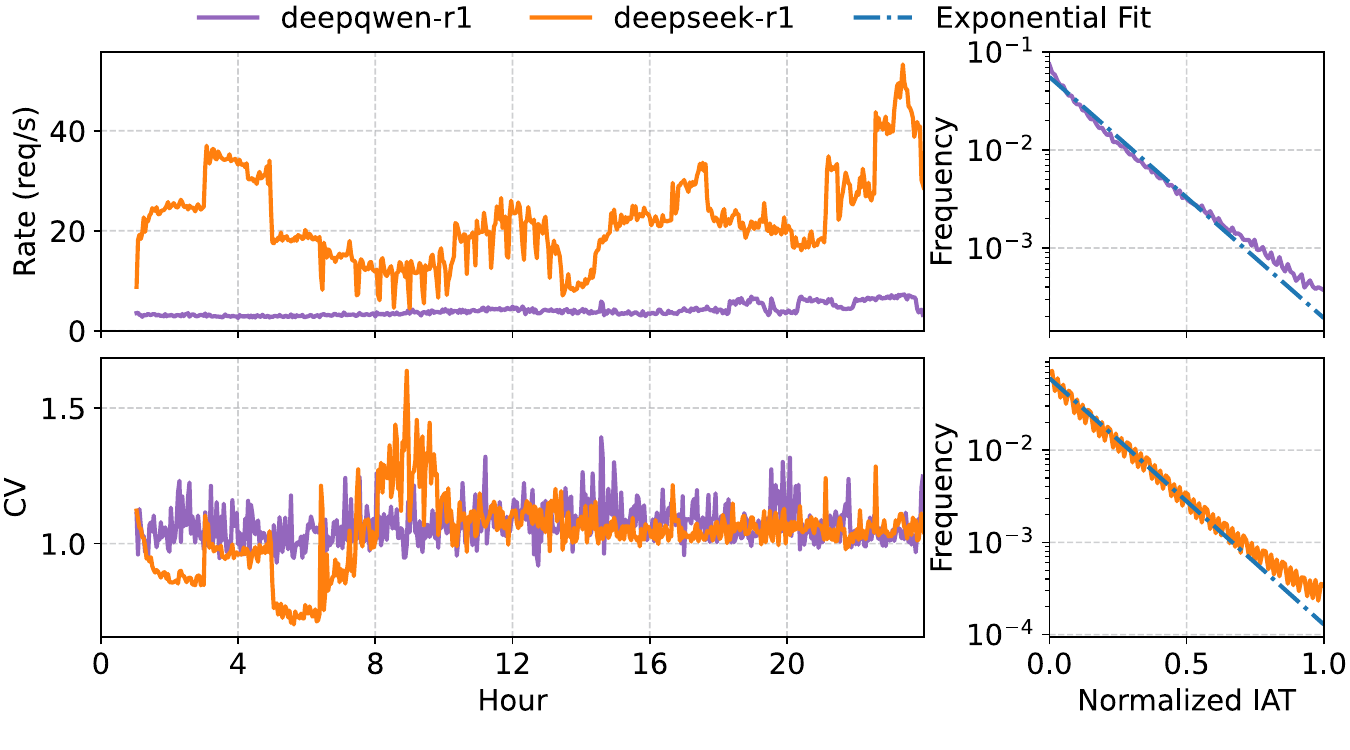}
    \vspace{-0.2in}
    \caption{
        Characterization of request arrival patterns in \ds and \dsqwen. 
        \emph{Left}: Rate and burstiness shifts over a day. 
        \emph{Right}: Normalized inter-arrival time distributions.
    }
    \label{fig:reasoning-arrival}
    \vspace{-0.1in}
\end{figure}

\begin{figure}[t!]
    \centering
    \begin{subfigure}[t]{0.45\linewidth}
        \includegraphics[width=\linewidth]{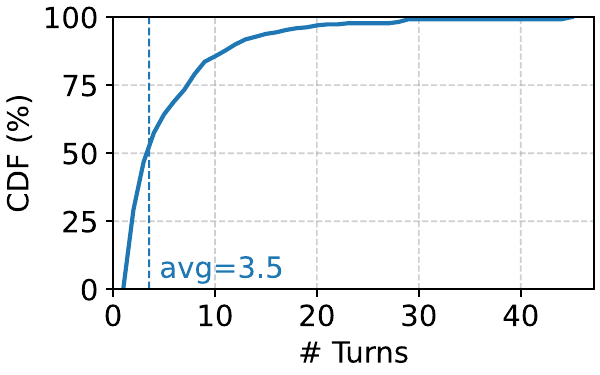}
        \vspace{-0.25in}
        \caption{CDF of conversation turns.}
        \label{fig:mt:cdf}
    \end{subfigure}
    \hfill
    \begin{subfigure}[t]{0.45\linewidth}
        \includegraphics[width=\linewidth]{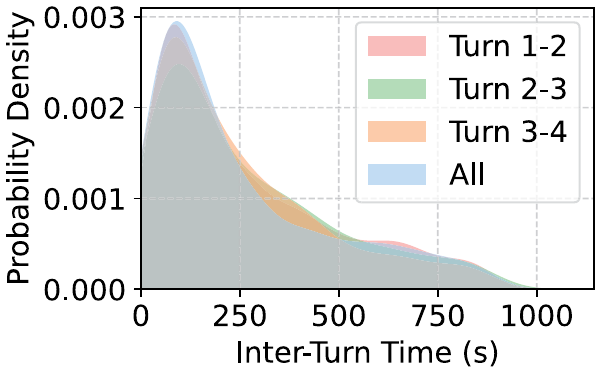}
        \vspace{-0.25in}
        \caption{PDF of inter-turn times.}
        \label{fig:mt:itt}
    \end{subfigure}
    \vspace{-0.05in}
    \caption{
        Characterization of conversations in \ds.
    }
    \label{fig:mt}
    \vspace{-0.15in}
\end{figure}

\stitle{Non-bursty arrivals.}
Figure~\ref{fig:reasoning-arrival} illustrates the arrival pattern for both 
\ds and \dsqwen over a day. On the left, the CV of request arrivals 
remains mostly close to or even less than 1 despite the diurnal rate shift, 
indicating that both workloads are non-bursty (especially compared with those 
in Figure~\ref{fig:rate-cv}). The right side of Figure~\ref{fig:reasoning-arrival} 
further validates this fact, showing that the Exponential distribution fits 
the inter-arrival time distribution quite well (\ie the arrival is roughly modeled 
by Poisson processes). 

\stitle{Characterizing multi-turn conversations.}
Holding multi-turn conversations is an essential capability of LLMs~\cite{wang2024mint, wu2023autogen}, 
and it also introduces a special pattern to request arrivals: 
intuitively, earlier requests foretell the \emph{reoccurrence} 
of follow-up conversations, which may alter the workload burstiness. 

We thus conduct a dedicated characterization of multi-turn requests found in \ds.
Within our 12-hour analysis window, we have identified\footnote{
    Our method is not accurate for many reasons: parts of conversations could fall 
    out of the analyzed window, or the messages could be altered by the log store. 
    Still, the resulting workload is reasonably large for analysis.
} 188,986 multi-turn 
requests out of 1,964,415 total requests, forming 57,205 conversations.
Figure~\ref{fig:mt:cdf} shows the distribution of the conversation lengths, averaging 3.5. 
The distribution of inter-turn time (ITT), \ie the time between the arrival of consecutive turns, 
is detailed in Figure~\ref{fig:mt:itt}. 
In general, ITTs concentrate around 100 seconds, with an extremely long tail 
(the figure is truncated at the $75_{\mathrm{th}}$ percentile for visualization). 

\begin{figure}[t!]
    \centering
    \includegraphics[width=.94\linewidth]{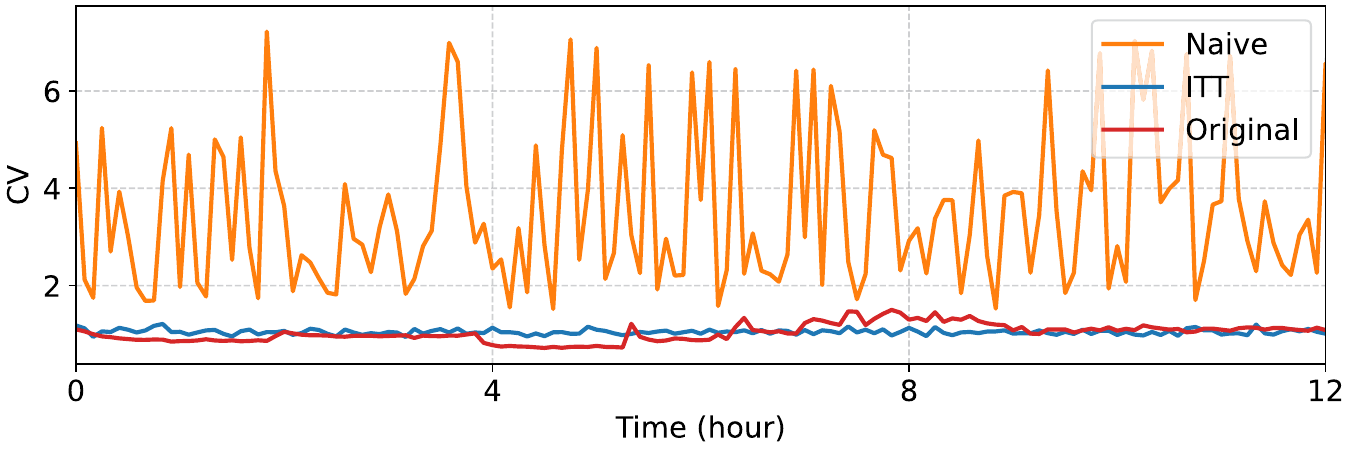}
    \vspace{-0.15in}
    \caption{
        Comparison of two upsampling methods for a workload containing only 
        multi-turn requests.
    }
    \label{fig:mt:upsample}
    \vspace{-0.15in}
\end{figure}

\stitle{Impact of multi-turn conversations.}
Since multi-turn requests constitute almost 10\% of the \ds workload, their pattern 
has a specific impact on workload characteristics. 
To demonstrate this, we apply two upsampling methods to the identified multi-turn 
requests, scaling them to the same size as the original workload. The \emph{Naive} 
method is agnostic about the conversations and simply scales the inter-arrival time, 
while the \emph{ITT} method works by scaling the arrival time between conversations, 
leaving the ITT distribution unchanged. 
Figure~\ref{fig:mt:upsample} compares the upsampled and original workloads 
by measuring the workload burstiness over time, highlighting a substantial difference: 
\emph{Naive} produces a highly bursty workload, while the \emph{ITT}-workload 
is even more stable than the original.
It is thus essential for realistic workloads to faithfully reflect the multi-turn 
conversation pattern by adhering to ITT distributions in Figure~\ref{fig:mt:itt}.

\begin{figure}[t!]
    \begin{subfigure}[t]{0.27\linewidth}
        \centering
        \includegraphics[width=\linewidth]{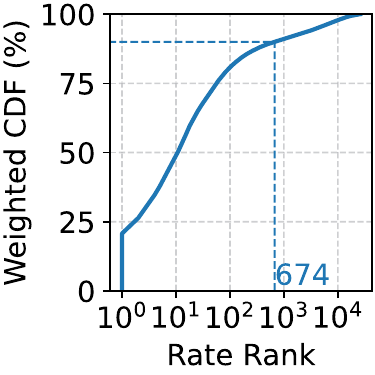}
        \vspace{-0.25in}
        \caption{}
        \label{fig:reasoning-client:rate}
    \end{subfigure}
    \hfill
    \begin{subfigure}[t]{0.27\linewidth}
        \centering
        \includegraphics[width=\linewidth]{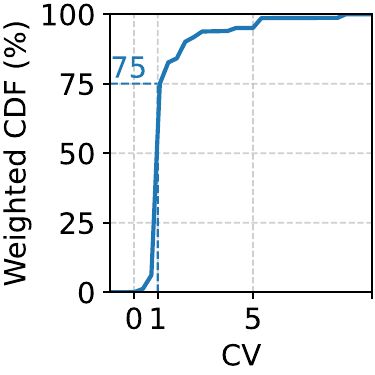}
        \vspace{-0.25in}
        \caption{}
        \label{fig:reasoning-client:cv}
    \end{subfigure}
    \hfill
    \begin{subfigure}[t]{0.44\linewidth}
        \centering
        \includegraphics[width=\linewidth]{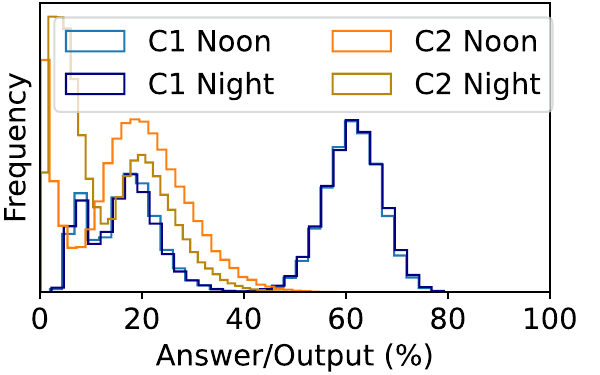}
        \vspace{-0.25in}
        \caption{}
        \label{fig:reasoning-client:ratio}
    \end{subfigure}
    \vspace{-0.15in}
    \caption{
        Client decomposition for \ds. 
        (a) weighted CDF of client arrival rate.
        (b) weighted CDF of client burstiness.
        (c) output length breakdown of top clients (C1 and C2).
    }
    \label{fig:reasoning-client}
    \vspace{-0.2in}
\end{figure}

{
\smallskip 
\begin{mdframed}[backgroundcolor=gray!10,linecolor=black,]
    \textbf{Finding~\finding{f:reason-arrival}}: Request arrival in reasoning workloads 
    is impacted by the reoccurring pattern of multi-turn conversations and 
    appears less bursty.
\end{mdframed}
}

\subsection{Decomposition of Reasoning Workloads}
\label{sec:reasoning:client}

Figure~\ref{fig:reasoning-client} presents client behaviors in the reasoning workload \ds. 
Interestingly, according to Figure~\ref{fig:reasoning-client:rate}, top clients in \ds 
are shown to be less substantial in comparison with other workloads 
(Figures~\ref{fig:hetero} and \ref{fig:mm:client-cdf}): out of 25,913 clients, the top 10 clients 
only constitute half of the requests. Furthermore, the proportion of non-bursty 
clients (Figure~\ref{fig:reasoning-client:cv}) is also significantly higher, likely 
contributing to the overall non-burstiness of the workload.
In addition, we observe again the bimodal distribution in the breakdown of request output lengths 
across multiple top clients, as depicted in Figure~\ref{fig:reasoning-client:ratio}. 
This implies that the pattern revealed in Figure~\ref{fig:reasoning-data:ratio} 
can still be causally modeled on a per-client basis, where the day-and-night shift 
of the answer length ratio is attributed to the fluctuation of client rates.

{
\smallskip 
\begin{mdframed}[backgroundcolor=gray!10,linecolor=black,]
    \textbf{Finding~\finding{f:reason-client}}: Clients in reasoning workloads 
    exhibit less skewed rates and less bursty arrivals, while also showing 
    the bimodal pattern in terms of data distributions.

\end{mdframed}
}

\section{Workload Generation}
\label{sec:reconstruct}

Motivated by the many findings in our characterization, we build \sysname, 
a principled framework for generating workloads that incorporate the 
realistic characteristics revealed in previous sections. 
Next, we describe our framework (\S\ref{sec:reconstruct:framework}), validate 
its accuracy (\S\ref{sec:reconstruct:accuracy}), and show its benefits 
for benchmarking serving systems with a real-world use case (\S\ref{sec:reconstruct:use}).

\subsection{\sysname Framework}
\label{sec:reconstruct:framework}

\begin{figure}[t!]
    \centering
    \includegraphics[width=.92\linewidth]{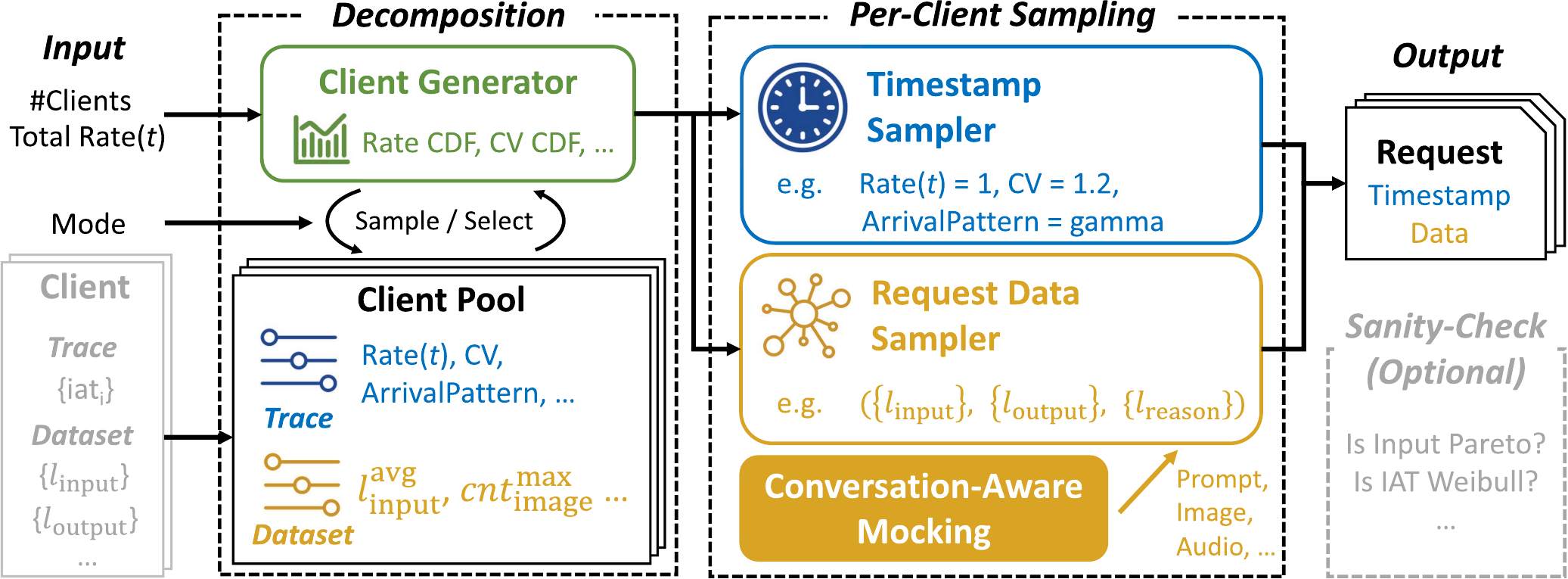}
    \vspace{-0.05in}
    \caption{
        Overview of the \sysname framework. 
        The color gray indicates optional requirements; 
        \eg users can still use \sysname without providing additional client information.
    }
    \label{fig:framework}
    \vspace{-0.14in}
\end{figure}

Figure~\ref{fig:framework} presents an overview of \sysname,  
which is centered around \emph{clients}. Essentially, \sysname samples requests 
on a \emph{per-client} basis, and aggregates them to compose realistic workloads.
Each client in \sysname is described by its trace and dataset, both 
of which can be either parameterized (\eg modeling a trace with the Gamma distribution) 
or provided as data samples (\eg a set of prompt lengths).

To use \sysname, a user starts by providing the total number of clients,
as well as a target total arrival rate. \sysname then relies on the 
{\tt Client Generator} to characterize each client, either by sampling from the 
{\tt Client Pool} pre-configured with realistic client behaviors, 
or by selecting from a set of user-specified clients with custom traces and 
datasets.
Next, \sysname samples the request timestamps and data for each client with 
the {\tt Timestamp Sampler} and {\tt Request Data Sampler}, scaling client 
rates according to the total rate and generating data via conversation-aware 
mocking to preserve shared conversation histories. 
Lastly, \sysname combines the timestamps and data to produce a final workload.

\begin{figure*}[t!]
    \centering
    \includegraphics[width=0.9\linewidth]{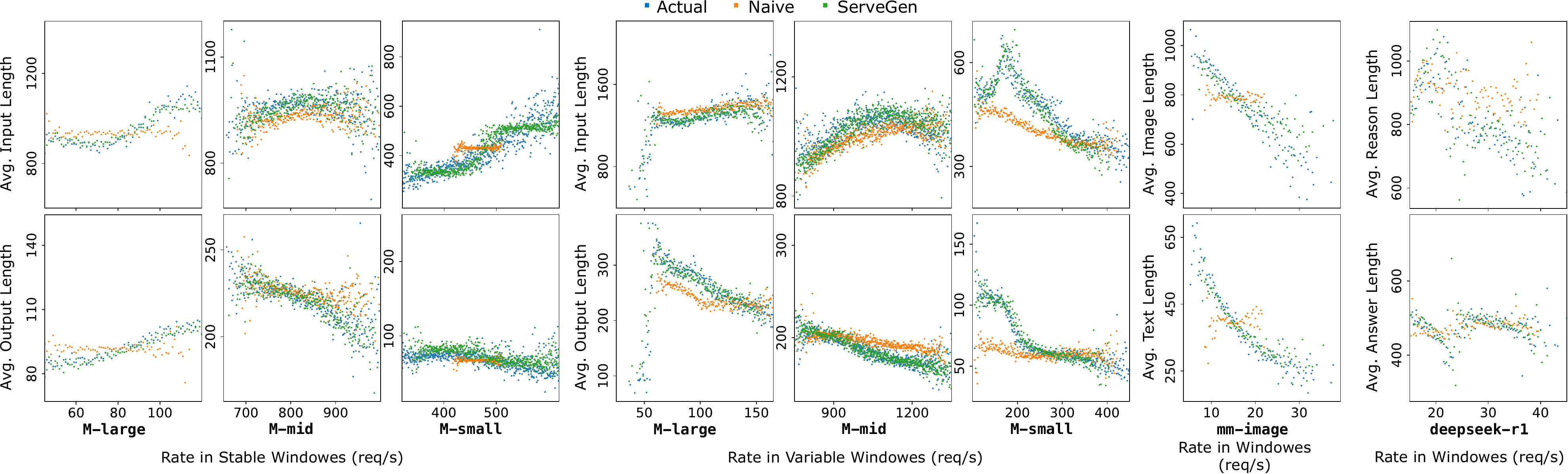}
    \vspace{-0.1in}
    \caption{
        Comparison of workload generation accuracy.
    }
    \label{fig:acc}
    \vspace{-0.15in}
\end{figure*}

\sysname utilizes the findings reported in previous sections to 
generate realistic workloads and ensure ease of use. 
For example, given Finding~\ref{f:ratecv}, the client rates and the total rate in 
\sysname are parameterized over the current time $t$, enabling workload generation 
with varying rates. 
Furthermore, we apply Finding~\ref{f:client} in the {\tt Client Generator} 
to produce heterogeneous clients (\ie sampling clients according to real 
rate and CV) and incorporate other findings on data distributions to configure 
the {\tt Client Pool} with parameterized real-world clients\footnote{
    For confidentiality, we release parameterized and sanitized data 
    instead of full data samples. Conversations are selectively released as hashed tokens. 
}. Users may optionally use Findings~\ref{f:iat} and \ref{f:io} 
to sanity-check the overall statistics of generated workloads.

\subsection{Generation Accuracy}
\label{sec:reconstruct:accuracy}

We validate that the per-client generation approach in \sysname captures the realistic 
characteristics of our workloads by measuring the generation accuracy with respect to Findings~\ref{f:ratecv}, \ref{f:ioshift}, 
\ref{f:mm-data}, and~\ref{f:reason-data}. Specifically, since the workload arrival patterns  
and data distributions undergo significant shifts over time, 
we aim to demonstrate that \sysname produces workloads that exhibit similar characteristics.

\stitle{Setups and metrics.}
We target the variability of data distributions in 3-hour time periods 
across raw workloads (Actual) and generated workloads that match the overall statistics 
(\eg length distributions for the full period). 
In each period, we measure the average values of relevant request data 
(\eg average input lengths for \qmax) in 3-second windows, and plot them against 
the request rates in those windows.
For language workloads, we explicitly differentiate 
between stable (\ie the request rate fluctuates around a certain value) and variable 
(\ie the overall request rate is rising or dropping) periods.
Intuitively, the shifting patterns in actual workloads should result in visible 
variability, which \sysname should be able to match with generated workloads.

\stitle{Configurations and baselines.}
We configure \sysname to select real clients and match the corresponding total rate, 
effectively resampling the workloads over client decomposition. 
In contrast, the baseline approach, referred to as \textsc{Naive}, directly resamples 
each workload as a whole to match the overall statistics with \sysname.  
\textsc{Naive} is meant to represent the workload generation method 
(\eg sampling ShareGPT over Poisson processes) used in many existing works~\cite{fastserve,shepherd,AlpaServe,wu2024loongserve} 
when direct workload replay is not an option because published workloads 
do not exist or mismatch with benchmarking setups.
For variable periods, the total rate in \textsc{Naive} is also parameterized 
by time to ensure fair comparisons.

\stitle{Results.}
Figure~\ref{fig:acc} displays the generation accuracy of the two approaches. 
In every case, the workload produced by \sysname is shown to be more realistic: 
the green plot (\sysname) matches the actual plot much better compared 
with \textsc{Naive}.

Furthermore, the results reveal two major drawbacks of the \textsc{Naive} workloads.
$(i)$ They can be less variable in terms of request rate, despite 
their overall burstiness. This is particularly evident during stable periods, where the 
blue and green scatter plots span considerably wider horizontally, indicating more extreme values 
for the arrival rate.
$(ii)$ They barely capture the correlation between rates and data distributions, 
which is non-trivial in real workloads (see the blue scatter plot). 
Such correlations are not surprising given our per-client characterization---large or small 
short-term rates are likely caused by bursty top clients, and the workload data 
distributions are expected to shift correspondingly toward or away from the 
client data distributions.

\subsection{Use Case \#1: Instance Provisioning}
\label{sec:reconstruct:use}

\begin{figure}[t!]
    \centering
    \begin{subfigure}{.378\linewidth}
      \centering
      \includegraphics[width=\linewidth]{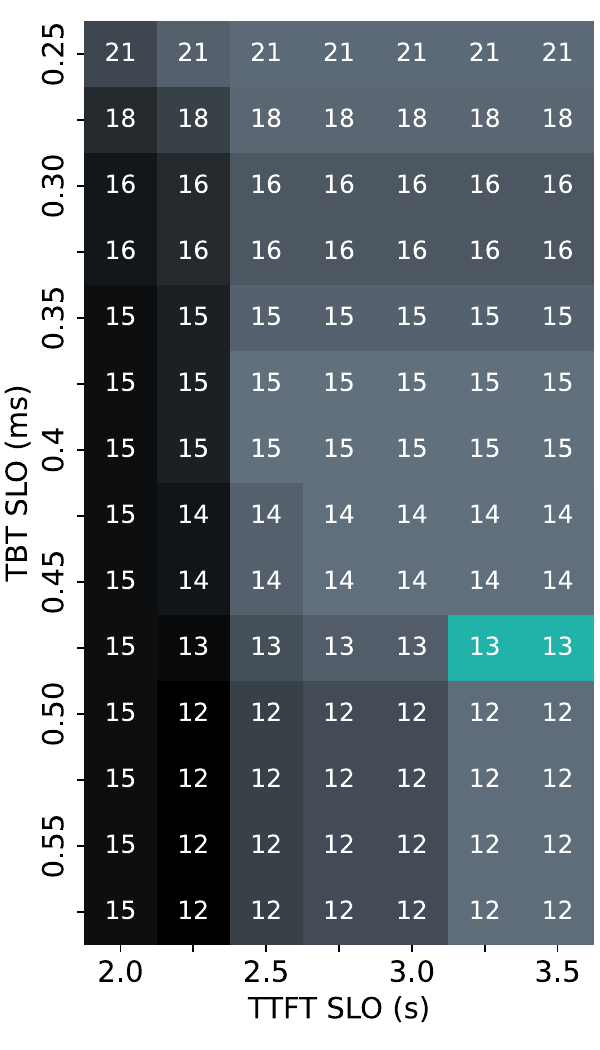}
      \vspace{-0.3in}
      \caption{\textsc{Naive} provisioning.}
      \label{fig:use:provision:naive}
    \end{subfigure}
    \begin{subfigure}{.4725\linewidth}
        \centering
        \includegraphics[width=\linewidth]{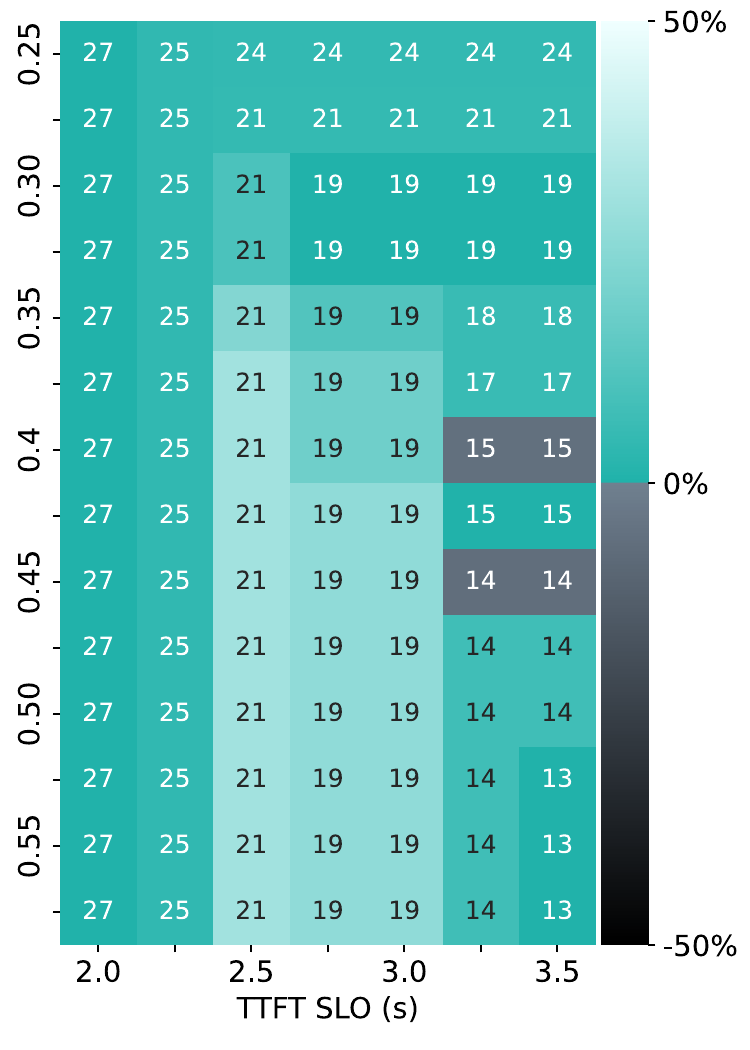}
        \vspace{-0.3in}
        \caption{\sysname provisioning.}
        \label{fig:use:provision:acc}
    \end{subfigure}

    \vspace{-0.1in}
    \caption{
        Provisioning results using the \textsc{Naive} approach and \sysname, 
        respectively. In each cell, the number indicates the provisioned instances, 
        while the color shows the over-provisioning percentage.
    }
    \label{fig:use:provision}
    \vspace{-0.2in}
\end{figure}

We now put \sysname to use, illustrating how it helps with benchmarking 
LLM serving systems by running the generated workloads on vLLM~\cite{vllm}, a 
representative LLM serving system with wide adoption. 
Particularly, we investigate an \emph{instance-provisioning} scenario,
\ie determining the minimum number of inference instances required to serve a workload 
while maintaining certain service-level objectives (SLOs). 
Next, we benchmark a vLLM instance with workloads produced by both \sysname and the 
\textsc{Naive} approach (as defined in \S\ref{sec:reconstruct:accuracy}) to obtain 
provisioning results, and then evaluate how well these results scale when 
serving real workloads. 

\stitle{Detailed setups.}
We select a 10-minute period of \qmax comprising 30,000 requests as the target workload, 
and set each instance to consist of 2 NVIDIA A100 (80GB) GPUs running a Qwen2.5-14B model\footnote{
    We opt for a smaller model than \qmax due to budget constraints.
} with pipeline parallelism~\cite{megatron,AlpaServe}. 
Next, for a grid of target time-to-first-token (TTFT) and time-between-token (TBT) SLOs, 
we benchmark one instance with workloads generated via both \sysname and \textsc{Naive}, 
adjusting the workload rate to find the maximum rate each instance can (supposedly) 
sustain without violating the SLOs (measured as P99 values), and thus derive the 
number of instances needed in each case. Lastly, we check the results by running 
the actual \qmax workload with the provisioned number of instances, recording the 
actual SLO delivered.

\stitle{Results.}
Figure~\ref{fig:use:provision} reports the provisioning results, where the number in 
each heatmap cell represents the provisioned instance count using either \textsc{Naive} or 
\sysname, and the cell color indicates the over- or under-provisioning percentage. 
For example, when the target P99 TTFT is 2.25s and TBT is 0.5s, \sysname results in  
provisioning 25 instances (4\% over the actual number needed), while \textsc{Naive} results in  
only 12 instances (50\% under-provisioning). Overall, Figure~\ref{fig:use:provision:naive} 
verifies that the \textsc{Naive} workloads are \emph{misleadingly easier to serve} than real 
workloads. Meanwhile, Figure~\ref{fig:use:provision:acc} fits the actual provisioning 
results much better, highlighting that the workloads generated by \sysname can 
better reflect the system performance in real-world deployment.

\begin{figure*}[t!]
    \centering
    \begin{subfigure}[t!]{.31\linewidth}
      \centering
      \includegraphics[width=\textwidth]{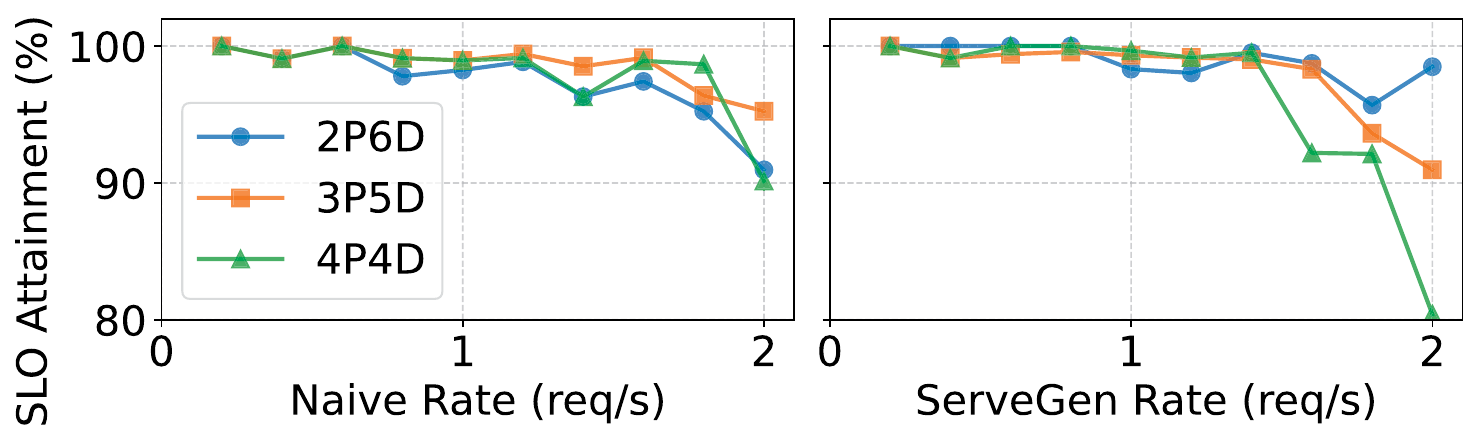}
      \vspace{-0.2in}
      \caption{Base SLO}
      \label{fig:use2:pd-base}
    \end{subfigure}
    \hfill
    \begin{subfigure}[t!]{.31\linewidth}
        \centering
        \includegraphics[width=\textwidth]{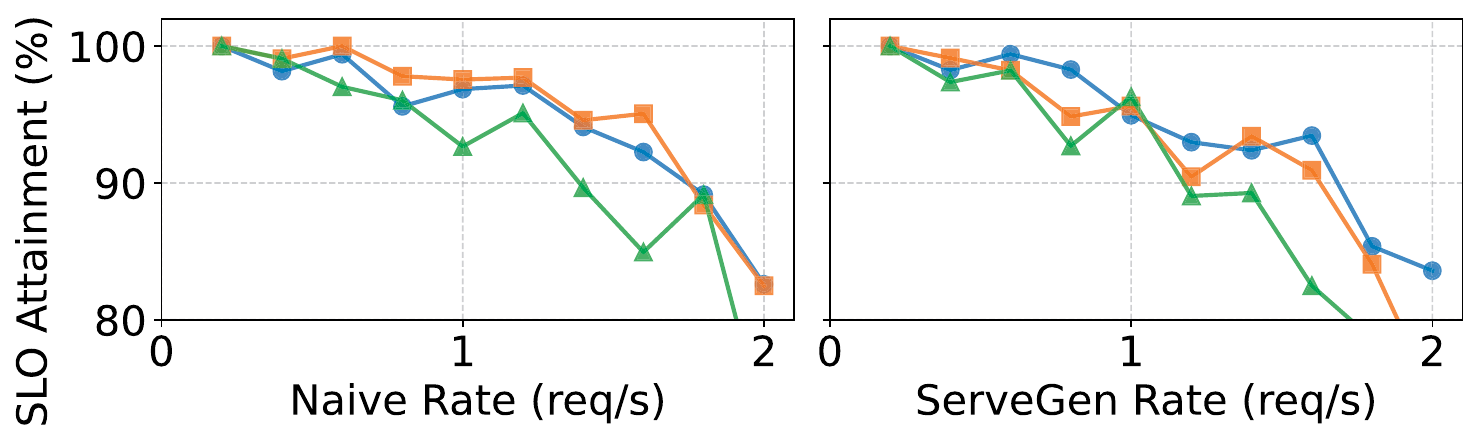}
        \vspace{-0.2in}
        \caption{Tight TBT SLO}
      \label{fig:use2:pd-tbt}
    \end{subfigure}
    \hfill
    \begin{subfigure}[t!]{.31\linewidth}
        \centering
        \includegraphics[width=\textwidth]{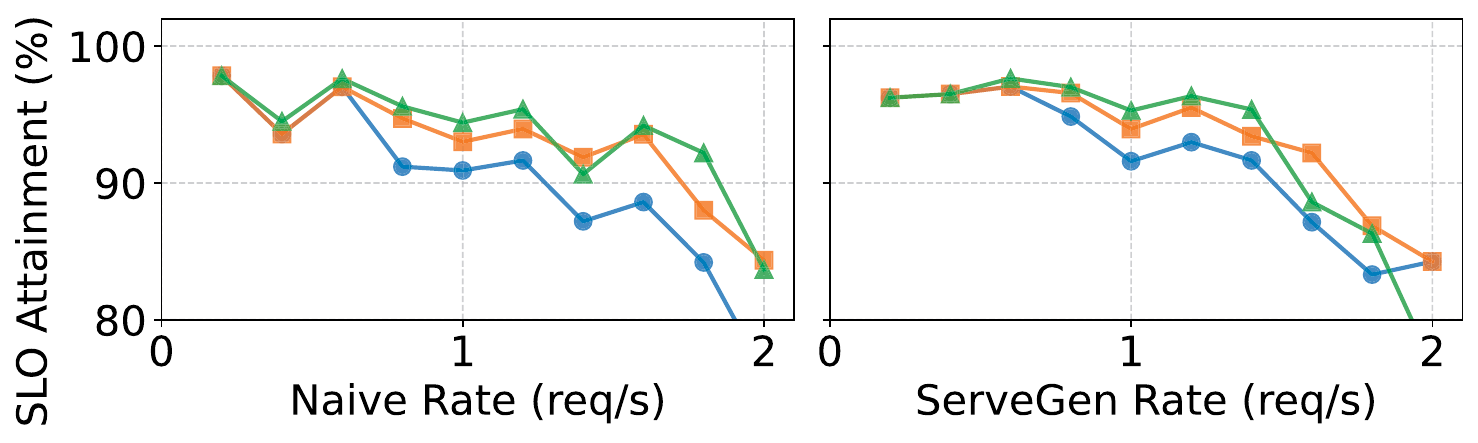}
        \vspace{-0.2in}
        \caption{Tight TTFT SLO}
      \label{fig:use2:pd-ttft}
    \end{subfigure}
    \vspace{-0.05in}
    \caption{
        SLO attainment under various setups.
        \textbf{Base} SLO: 8s TTFT, 60ms TBT.
        \textbf{Tight} SLO: 4s TTFT, 30ms TBT (half of \textbf{Base}).
    }
    \label{fig:use2}
    \vspace{-0.2in}
\end{figure*}

\subsection{Use Case \#2: PD-Disaggregation}
\label{sec:reconstruct:use2}

We also demonstrate how \sysname is beneficial to understanding the impact of serving system 
design choices with a case study on PD-disaggregation~\cite{distserve}, an architectural 
innovation that is widely adopted in both academia~\cite{splitwise,pdserve,semipd} 
and industry~\cite{mooncake,deepseek-inference,sglang-deploy,dynamo}.
While it disaggregates the prefill and decoding phases to avoid interference and achieve tailored 
optimization, recent research~\cite{buzz,arrow} indicate that it can be tricky to fully characterize  
its performance impact in real-world deployments.
Here, we show with \sysname that the lack of proper workload understanding is largely at fault.

\stitle{Detailed setups.}
We follow the same steps in \S\ref{sec:reconstruct:use} to generate workloads via 
\sysname and \textsc{Naive} (with identical overall rate and data distribution). 
Each workload is then used to benchmark a Qwen2.5-72B model deployed on 4 nodes with 8 H20 
GPUs each, running PD-disaggregated SGLang~\cite{sglang} (tensor parallelism~\cite{megatron} set to 4). 
We vary the prefill and decoding resource scaling (\eg 3P5D indicates 3 prefill instances and 
5 decoding instances) to examine the best configuration.

\stitle{Results.}
Figure~\ref{fig:use2} shows the benchmarking results under different target SLO setups, 
where PD-disaggregation displays highly workload-sensitive performance. 
Most interestingly, the results may disagree about the best PD configuration under \textsc{Naive} 
and \sysname workloads even under similar distributions. 
In two of the three cases (Figures~\ref{fig:use2:pd-base} and~\ref{fig:use2:pd-tbt}), the best configuration 
under \sysname is 2P6D, which happens to be a worse configuration reported by \textsc{Naive} 
benchmarking. This discrepancy is likely due to more intricate patterns of long-tailed requests 
in \sysname, leading to more extreme load-imbalance and a greater demand for decoding instances.

In practice, Figure~\ref{fig:use2} may manifest itself as "unpredictable" performance drops  
where PD-disaggregation delivers suboptimal performance due to transient load-imbalance as 
workloads shift (characterized by \sysname instead of \textsc{Naive}), 
highlighting the need for new design explorations (\eg dynamic load-balancing, explicit performance modeling) 
to further improve the PD-disaggregation approach.
\section{Discussion}
\label{sec:discussion}

\stitle{Fostering future research.}
The aforementioned findings have already benefited several development teams at Alibaba Cloud Model Studio, 
including those focused on inference engine optimization, resource planning, 
and request scheduling. Meanwhile, \sysname can guide further research on LLM serving 
in many other ways, and we point out two potential directions here.
First, our multimodal workload analysis reveals that a significant portion of 
TTFT stems from preprocessing (\ie downloading, normalization, and encoding). 
This highlights the importance of conducting full-stack optimizations (\eg decoupling 
the modality encoders and scaling them independently according to Finding~\ref{f:mm-data}, 
rather than focusing solely on the LLM). 
Second, our analysis of multi-turn conversations in reasoning workloads reveals 
that the arrival pattern for these requests is non-bursty (Finding~\ref{f:reason-arrival}), 
providing valuable insights for improving short-term workload predictability in 
conversational scenarios.

\stitle{Limitations of \sysname.} 
While \sysname covers mainstream LLM serving workloads, there are several aspects  
that require further study. First, some complex LLM applications adopt
\emph{plugin calls}, where a series of functions are called prior to model
inference, performing web searches, database queries, or calling external
APIs. The dependent execution of different functions collectively determines the
end-to-end execution time, and the output length is influenced as well. We leave
characterizing LLM serving with plugin calls as an important area for future work.
Second, \emph{prefix caching}~\cite{sglang} enables sharing intermediate KV cache
between requests with common prompt prefixes. However, characterizing prefix
caching requires access to the content of requests which we currently opt out of 
due to confidentiality obligations. Third, our analysis of individual workloads 
does not yet span long enough for \emph{longitudinal} studies (\eg trends in serving workloads 
over months).
\section{Related Work}
\label{sec:related}

\stitle{LLM serving analysis.}
Prior work has analyzed many aspects of LLM serving to faciliate optimization and development.
Specific to LLM serving \emph{workloads}, BurstGPT~\cite{burstgpt} and
LMM~\cite{LMM} have characterized language and image-to-text model serving
workloads, while a series of other studies have performed brief analyses  
from certain viewpoints such as burstiness~\cite{muxserve},
computational load~\cite{distserve,splitwise}, prefix-sharing~\cite{mooncake,cao2025locality},
and energy efficiency~\cite{DynamoLLM}. 
On the other hand, dedicated simulators (\eg AstraSim~\cite{astrasim2}, Vidur~\cite{vidur}) have been built to 
tackle high-fidelity simulation of serving \emph{systems} for performance analysis. 
\sysname serves as a comprehensive characterization of LLM serving workloads 
with a larger scale and scope, while also strictly complementing inference simulation and 
hopefully unlocking realistic end-to-end performance modeling.

\stitle{Workload modeling and generation.} 
Existing research has characterized various workloads in alternative machine learning 
scenarios such as GPU deep learning~\cite{PAI-train,PAI,Philly,Helios} and LLM development~\cite{Acme}, 
providing many valuable insights.
Some works~\cite{syn1,syn2,cloud1,tracegen,PAI} have also proposed generating realistic workloads 
by modeling them with historical data, yet they mostly target generic cloud-computing workloads. 
BurstGPT~\cite{burstgpt} is a recent work that tackles workload modeling for LLM serving, 
which uses a parameterized Gamma process to account for variant burstiness in LLM serving. 
Meanwhile, a large body of prior research~\cite{fastserve,shepherd,AlpaServe,wu2024loongserve} 
on LLM serving has relied on the \textsc{Naive} approach and simply combined traces and datasets.
We hope the release of \sysname can foster LLM serving research by covering 
multiple workload categories and modeling them more accurately with client 
decomposition, while ensuring ease of use for practitioners.

\section{Conclusion}
\label{sec:conclusion}

We present a comprehensive characterization of real-world LLM serving workloads for
language, multimodal, and reasoning models. We unveil various characteristics and
summarize meaningful findings. Based on these findings, we provide \sysname, a
principled framework that generates realistic serving workloads by composing 
them on a per-client basis. 
We demonstrate the benefits of \sysname via case studies on instance provisioning 
and PD disaggregation.

\stitle{Acknowledgments.}
We thank our shepherd, Ramachandran Ramjee, and the anonymous reviewers for their insightful feedback.
This work was supported in part by the Scientific Research Innovation Capability Support Project for Young Faculty 
under Grant ZYGXQNJSKYCXNLZCXM-I1. Xin Jin and Xue Li are the corresponding authors. 
Yuxing Xiang, and Xin Jin are also with the Key Laboratory of High Confidence Software 
Technologies (Peking University), Ministry of Education.

\clearpage

\balance
\bibliographystyle{plain}
\bibliography{paper}

\end{document}